\documentclass[aip,jcp]{revtex4-1}
\usepackage{graphicx}
\usepackage{natbib}
\usepackage{bm}
\usepackage{amsmath}
\usepackage{amssymb}
\usepackage{subfigure}

\newtheorem{lemma}{Lemma}

\newcommand{\vnorm}[1]{\|#1\|}
\newcommand{\mcal}{\mathcal}
\newcommand{\divr}{\operatorname{div}}

\newcommand{\stress}{\bm{\sigma}}
\newcommand{\pot}{\mathcal{V}}
\newcommand{\eref}[1]{(\ref{#1})}
\newcommand{\sref}[1]{Section~\ref{#1}}
\newcommand{\fref}[1]{Fig.~\ref{#1}}
\newcommand{\real}[1]{\mathbb{R}^#1}

\begin{document}


\title{Stress and heat flux for arbitrary multi-body potentials: A unified framework}



\author{Nikhil~Chandra~Admal}
\email[]{admal002@umn.edu}
\author{E.~B.~Tadmor}
\email[]{tadmor@aem.umn.edu}
\affiliation{Department of Aerospace Engineering and Mechanics, The University of Minnesota, Minneapolis, Minnesota 55455}


\date{\today}

\begin{abstract}
A two-step \emph{unified framework} for the evaluation of continuum field 
expressions from molecular simulations for arbitrary interatomic potentials
is presented. First, pointwise continuum fields are obtained using a 
generalization of the Irving--Kirkwood procedure to arbitrary multi-body potentials.
Two ambiguities associated with the original Irving--Kirkwood procedure (which
was limited to pair potential interactions) are addressed in its generalization.
The first ambiguity is due to the non-uniqueness of the decomposition of the force 
on an atom as a sum of central forces, which is a result of the non-uniqueness of 
the potential energy representation in terms of distances between the particles.
This is in turn related to the \emph{shape space} of the system. The
second ambiguity is due to the non-uniqueness of the energy
decomposition between particles. The latter can be completely avoided through
an alternate derivation for the energy balance.  It is found that the expressions 
for the specific internal energy and the heat flux obtained through the 
alternate derivation are quite different from the original Irving--Kirkwood
procedure and appear to be more physically reasonable. Next, in the second step of the
unified framework, spatial averaging is applied to the pointwise field to obtain
the corresponding macroscopic quantities. These lead to expressions suitable
for computation in molecular dynamics simulations. It is shown that the important
commonly-used microscopic definitions for the stress tensor and heat flux vector 
are recovered in this process as special cases (generalized to arbitrary 
multi-body potentials). Several numerical experiments are
conducted to compare the new expression for the specific internal energy with
the original one.
\end{abstract}

\pacs{02.70.Ns, 05.20.-y, 45.50.-j, 46.15.-x}

\maketitle

\section{Introduction}
The idea of defining continuum fields from particle mechanics (for the
special case of pair potential interactions) was pioneered in
the landmark paper of Irving and Kirkwood\cite{ik1950}. Irving and Kirkwood
derived the equations of hydrodynamics from the principles of non-equilibrium
classical statistical mechanics and in the process established pointwise
definitions for various continuum fields. Under this procedure, basic
continuum fields including the
mass density, momentum density and the specific internal energy are defined \emph{a
priori} using
a probability density function. Using these definitions, expressions for the stress tensor and
the heat flux vector fields are obtained that identically satisfy the balance laws of continuum mechanics. 
The continuum fields obtained in Irving and Kirkwood's original paper\cite{ik1950}
involved a series expansion of the Dirac delta distribution, which is not
mathematically rigorous.\footnote{The derivation is non-rigorous in the sense that
    expressing the stress tensor as a series expansion is only possible when the
    probability density function, which is used in the derivation, is an
    analytic function of the spatial variables (see Ref\cite{noll1955}).}
In a follow-up study, Noll\cite{noll1955,lehoucq2010} proved
two lemmas, which allowed him to avoid the
use of the delta distribution and to obtain closed-form analytical
expressions for the continuum fields.

Since the Irving--Kirkwood procedure is stochastic in nature,
many problems arise when one tries to use the resulting expressions for a practical
calculation --- a key one being our lack of knowledge of the probability density
function. To avoid these difficulties, Hardy\cite{hardy1982} and independently
Murdoch\cite{murdoch1993,murdoch1994,murdoch2003,murdoch2007}
developed a simpler spatial averaging procedure that avoids the mathematical complexity of the Irving--Kirkwood
procedure. We refer to the procedure due to Hardy as the \emph{Hardy procedure}
and that due to Murdoch as the \emph{Murdoch procedure}.\footnote{Although the
Hardy and Murdoch procedures seem similar, they are notably
different in the derivation of the energy balance equation, and the resulting expressions for
energy density and the heat flux vector.} In these 
procedures, continuum fields are defined as direct spatial
averages of the discrete equations of motion using a normalized weighting
function. This approach also leads to a set of definitions
that identically satisfy the balance equations. Therefore, we have three different approaches for defining the
continuum fields from particle mechanics --- although originally developed for pair
potentials only.

Of the continuum fields, the
stress tensor has been studied most extensively.
In addition to the definitions for the stress tensor obtained from
the systematic approaches described above, a number of 
other definitions have been proposed in
the past dating back to the work of Cauchy\cite{cauchy1828a,cauchy1828b}
on the stress vector and Clausius\cite{clausius1870} on the virial 
stress.\footnote{See Ref\cite{admal2010} for a more detailed
historical review.} Efforts at obtaining microscopic definitions for the 
stress tensor (as well as other continuum variables) are ongoing; see for
example Refs\cite{tsai1979,dahler1995,cormier2001,zim2004,heinz2005,delph2005,morante2006,chen2006,mandadapu2009a,mandadapu2009b,thompson2009,rossi2010} for some important contributions. 
A recent article\cite{admal2010} by the authors extensively studies the definition for the stress tensor within a unified framework based on a generalization of the 
Irving--Kirkwood procedure to arbitrary multi-body potentials followed by a 
process of spatial averaging.  Through this unified framework
it is shown that all existing definitions, 
including the virial stress tensor,\cite{clausius1870} Hardy
stress tensor,\cite{hardy1982}  and Cauchy/Tsai stress tensor,\cite{cauchy1828a,cauchy1828b,tsai1979} 
which all seem to be derived from disparate approaches, follow as special cases
from a single stress expression. 
Furthermore, the derivation in Ref\cite{admal2010} reveals the subtle 
(and hitherto unrecognized issue) that interatomic potentials constitute
\emph{continuously differentiable extensions} to functions defined over a 
more limited domain. This is a vital part of the derivation with important implications for the uniqueness of the microscopic stress tensor --- an 
issue which is widely discussed in the literature cited above. Although there
have been a number of attempts to generalize the Irving--Kirkwood procedure and the
Hardy procedure to multi-body potentials (see
Refs\cite{zim2008,chen2006,zhang2004,heinz2005}), these attempts are either
restricted to specific potentials (see Refs\cite{chen2006,zhang2004}) or the source of non-uniqueness of the
stress tensor is not explicitly identified. In contrast,
the unified framework developed in Ref\cite{admal2010} applies
to \emph{arbitrary} multi-body potentials and rigorously characterizes the non-uniqueness of the
stress tensor.

\newcounter{fnnumber}
The aim of this paper is to continue to use this unified framework to study the energy
balance equation of continuum mechanics in the context of multi-body
potentials.  As noted earlier, in the original Irving--Kirkwood and the Hardy procedure, the 
definition for the potential part of the specific internal energy 
(for the special case of pair potentials in a mono-atomic system)
is assumed \emph{a priori} and the expression for the heat flux vector
is then derived to ensure that the energy balance equation is identically satisfied.
Unfortunately, this approach does not generalize to arbitrary multi-body potentials
(or even pair potentials with multiple species types) since it involves an ambiguous definition for the ``energy of an atom''. To the
best of the authors' knowledge, all the existing works (see
Refs\cite{zim2008,chen2006,zhang2004,marechal1983,torii2008}) which attempt to derive a
microscopic definition for the heat flux in the case of multi-body potentials by
generalizing the Irving--Kirkwood procedure or the Hardy procedure suffer from this ambiguity. For
example, in Ref\cite{chen2006} it was assumed that the energy corresponding to a
cluster of three particles interacting through a three-body potential is evenly
distributed among the particles. However, there is no symmetry argument to justify
this assumption.\footnote{The symmetry argument for equally
dividing the energy among particles only holds for identical atoms interacting
via a pair potential. It is lost for multi-species systems and for all higher-order potentials. For example, 
for a three-body potential (even in the case of a single-species system), the symmetry
between atoms is lost in all clusters of three particles which do not form an
equilateral triangle. The same reasoning applies for potentials of higher
order.}\setcounter{fnnumber}{\thefootnote} Furthermore,
even for the case of pair potential interactions, the original Irving--Kirkwood
approach leads to an expression for the heat flux vector which is not invariant
with respect to the addition of a constant to the potential energy of the system,
which is not physically reasonable. In contrast, in the Murdoch procedure,
the specific internal energy and heat flux vector are obtained together as part
of the derivation and the resulting expressions are consistent with physical
expectations.  Motivated by this, in this paper, we reformulate the Irving--Kirkwood
procedure using the method followed by Murdoch\cite{murdoch1994}. This approach leads to physically-acceptable expressions 
for the internal energy density and heat flux vector which are grounded in rigorous 
statistical mechanics principles and which does not require any energy
decomposition between the particles.  Furthermore, as noted above, our derivation 
extends those of Irving--Kirkwood and Murdoch to arbitrary multi-body 
potentials. Finally, the application of the spatial averaging step in the unified
procedure leads to expressions suitable for use in molecular dynamics
simulations. These expressions are compared with those from the original
Irving--Kirkwood formulation through a number of simple numerical experiments.

The following notation is used in this paper. Vectors are denoted by lower case
letters in bold font, while tensors of higher order are denoted by capital
letters in bold font. The inner product of two vectors is given by a dot ``$\cdot$'',
and their tensor product is given by the symbol ``$\otimes$''. The
inner product of two second-order tensors is denoted by ``:''. The gradient of a
vector field, $\bm v(\bm x)$, is denoted by $\nabla_{\bm x} \bm v(\bm x)$. A
second-order tensor, $\bm T$, operating on a vector, $\bm v$, is denoted by $\bm
T\bm v$. The divergence of a tensor field, $\bm T(\bm x)$, is denoted by
$\divr_{\bm x} \bm T(\bm x)$, which corresponds to $\partial \bm
T_{ij}/\partial \bm x_j$ in indicial notation (with Einstein's summation
convention).
\label{sec:intro}

\section{Continuum fields as phase averages}
\label{ch:phase}
Consider a system modeled as a collection of $N$ point particles,
each particle identified by an index $\alpha$ $(\alpha=1,2,\ldots,N)$. The position,
mass, and velocity of particle $\alpha$ are denoted by $\bm{x}_\alpha$,
$m_\alpha$ and $\bm{v}_\alpha$, respectively. We assume that the particles
interact through a continuously differentiable function $\pot(\bm x_1,\dots,\bm
x_N)$, which is called the \emph{potential energy} of the system. The complete microscopic state of
the system  at any instant of time is known from the knowledge of position and
velocity of each particle in $\real{3}$. Hence, the state of the system  at time
$t$ may be represented by a point in a $6N$-dimensional phase
space.\footnote{The usual convention is to represent the phase space via
positions and momenta of the particles. For convenience, in this section, we
instead use positions and velocities.} Let
$\Gamma$ denote the phase space. Therefore any point in $\Gamma$, can be represented as,
\begin{align}
(\bm{x}(t);\bm{v}(t)) :=
(\bm{x}_1(t),\dots,\bm{x}_N(t);\bm{v}_1(t),\dots,\bm{v}_N(t)).
\label{eqn:define_X}
\end{align}
In reality, the microscopic state of the system is never known to us and the
only observables identified are the macroscopic fields as defined in continuum
mechanics. We identify the continuum fields with macroscopic observables
obtained in a two-step process: (1) a pointwise field is obtained as a
statistical mechanics phase average; (2) a macroscopic field is obtained as a
spatial average over the pointwise field.  The phase averaging in step (1) is
done with respect to a continuously differentiable\footnote{The assumption that
the probability density function exists and it is continuously differentiable
can be considerably weakened by viewing $W$ as a generalized
function/distribution in the sense of Schwartz. For the sake of brevity we do not take this approach,
however, we later use a generalized function/distribution as a candidate for $W$
to arrive at expressions for continuum fields that can be used in a molecular
dynamics simulation. See \sref{sec:md} for details.} probability density function
$W:\Gamma \times \mathbb{R}^+ \rightarrow \mathbb{R}^+$ defined on all phase
space for all $t$. The explicit dependence of $W$ on time $t$, indicates that
our system need not be in thermodynamic equilibrium.

The basic idea behind the original Irving and Kirkwood procedure is to
prescribe the mass density, velocity and the specific internal energy fields,
which we call the input fields, and
derive the body force vector, stress tensor and the heat flux vector fields, which we call the
output fields, such that all the definitions are consistent with the balance
laws of mass, momentum and energy:  
\begin{align}
\label{eqn:fields}
\textbf{Input fields}  \quad \qquad & \qquad \textbf{Output fields} \notag \\
\left \{ \begin{array}{l} 
\mbox{mass density} \\
\mbox{velocity}\\
\mbox{specific internal energy} \end{array} \right \} 
&\rightarrow
\left \{ \begin{array}{l} 
\mbox{body force} \\
\mbox{stress} \\
\mbox{heat flux} \end{array} \right \}.
\end{align}
To arrive at
these definitions, we repeatedly use the following result of Liouville's
theorem, which describes the evolution of the probability density function: 
\begin{equation}
\label{eqn:liouville}
\frac{\partial W}{\partial t} + \sum_{\alpha=1}^{N} \left [ \bm{v}_\alpha \cdot \nabla_{\bm{x}_\alpha}W + \dot{\bm{v}}_\alpha \cdot \nabla_{\bm{v}_\alpha}W \right ] = 0.
\end{equation}
Since the force on a particle $\alpha$ is given by
\begin{equation}
\label{eqn:def_fi}
\bm{f}_\alpha := -\nabla_{\bm{x}_\alpha} \pot,
\end{equation}
equation \eref{eqn:liouville} can be rewritten as
\begin{equation}
\frac{\partial W}{\partial t} + \sum_{\alpha=1}^{N} \left [ \bm{v}_\alpha \cdot \nabla_{\bm{x}_\alpha}W - \frac{\nabla_{\bm{x}_\alpha} \pot }{m_\alpha} \cdot \nabla_{\bm{v}_\alpha}W \right ] = 0,
\label{eqn:useful_liouville}
\end{equation}
where, as stated before, $\pot(\bm{x}_1,\bm{x}_2,\dots,\bm{x}_N)$ denotes the potential
energy of the system. Equation~\eref{eqn:useful_liouville} is called
\emph{Liouville's equation}. 

\medskip
To proceed, we divide the potential energy into two parts:
\begin{enumerate}
\item An \emph{external} part, $\pot_{\rm ext}$, associated with long-range
interactions such as gravity or electromagnetic fields,
\item An \emph{internal} part, $\pot_{\rm int}$, associated with short-range
particle interactions. In general, the internal part of the potential energy is
also called the \emph{interatomic potential energy}. 
\end{enumerate}
We next define the input fields used in the Irving--Kirkwood procedure.

\subsection{Phase averaging}
\label{sec:phase}
Under the Irving--Kirkwood procedure, pointwise fields are defined as
phase averages. For example, the pointwise mass density field is defined as
\begin{equation}
\rho(\bm{x},t) := \sum_{\alpha} m_\alpha \int_{\real{{3N}} \times \real{{3N}}} W \delta (\bm{x}_\alpha - \bm{x}) \, d\bm{x} d\bm{v},
\label{eqn:define_density_delta}
\end{equation}
$\delta$ denotes the Dirac delta distribution, and $\sum_\alpha$ denotes summation from
$\alpha=1$ to $N$.  To avoid the Dirac delta distribution and for greater
clarity we adopt the notation introduced by Noll. Hence \eref{eqn:define_density_delta} can be rewritten as
\begin{align}
\rho(\bm{x},t) &= \sum_{\alpha} m_\alpha \int W \, d\bm{x}_1 \dots d\bm{x}_{\alpha-1} d\bm{x}_{\alpha+1}\dots d\bm{x}_N d\bm{v} \notag \\
&=: \sum_{\alpha} m_\alpha \left \langle W \mid \bm{x}_\alpha = \bm{x} \right \rangle,
\label{eqn:define_density}
\end{align}
where $\left \langle W \mid \bm{x}_\alpha = \bm{x} \right \rangle$ denotes an integral of $W$ over all its arguments except $\bm{x}_\alpha$, and $\bm{x}_\alpha$ is substituted with $\bm{x}$.

The second input field, which is the pointwise velocity field, is defined via the momentum density field, $\bm{p}(\bm{x},t)$, as follows:
\begin{align}
\bm{p}(\bm{x},t) &:= \sum_{\alpha} m_\alpha \left \langle W \bm{v}_\alpha \mid \bm{x}_\alpha = \bm{x} \right \rangle,\label{eqn:define_mom_density}\\ 
\bm{v}(\bm{x},t) &:= \frac{\bm{p}(\bm{x},t)}{\rho(\bm{x},t)}. \label{eqn:define_velocity}
\end{align}
The third input field, which is the specific internal energy, depends on the interatomic
potential. At this point, it must be noted that the
original Irving--Kirkwood procedure was limited to systems
interacting through a pair potential function:
\begin{align}
\label{eqn:pairpot}
\pot_{\rm int} &= \pot_{\rm int}(r_{12},\dots,r_{1N},r_{23},\dots,r_{(N-1)N})
\notag \\
&= \sum_\alpha \pot_\alpha,
\end{align}
where $\pot_\alpha$ is the energy of particle $\alpha$, defined as 
\begin{align}
\label{eqn:def_pot_alpha}
\pot_\alpha :=
\frac{1}{2}\Bigg[
\sum_{\substack{\beta \\ \beta < \alpha}}
    \phi_{\beta\alpha}(r_{\beta\alpha}) + 
\sum_{\substack{\beta \\ \beta > \alpha}}
    \phi_{\alpha\beta}(r_{\alpha\beta})
\Bigg].
\end{align}
and $\phi_{\alpha\beta}$ ($\alpha < \beta$) is the energy corresponding to the
interaction of the pair $(\alpha,\beta)$. In this case, the specific internal energy is defined as
\begin{align}
\label{eqn:ener}
\epsilon(\bm x,t) := \epsilon_{\rm k}(\bm x,t) + \epsilon_{\rm v}(\bm x,t),
\end{align}
where
\begin{align}
\label{eqn:ener_kin}
\rho \epsilon_{\rm k}(\bm x,t) := \frac{1}{2}\sum_\alpha m_\alpha \langle \vnorm{\bm
v_\alpha}^2 W \mid \bm x_\alpha = \bm x \rangle,
\end{align}
is the kinetic contribution to the specific internal energy , and
\begin{align}
\label{eqn:ener_pot}
\rho \epsilon_{\rm v}(\bm x,t) := \sum_\alpha 
\langle \pot_\alpha W \mid \bm x_\alpha = \bm x \rangle,
\end{align}
is the potential contribution to the specific internal energy. According to the
definition given in \eref{eqn:ener}, the specific internal energy at $(\bm x,t)$
is the weighted sum of the energy of each particle with the probability that it
is at $\bm{x}$ at time $t$. It is clear from the definition in \eref{eqn:def_pot_alpha} that the
interaction energy $\phi_{\alpha\beta}$, between any two particles $\alpha$ and
$\beta$, is shared equally between the particles $\alpha$ and $\beta$. This is
plausible for systems with identical particles interacting with pair potential,
but there is no \emph{a priori} physically motivated way of deciding how to distribute
the energy for systems interacting through a multi-body potential. This is one
of the primary reasons why the definition for the specific internal energy and
the energy balance equation has to be re-examined as we do later in
\sref{sec:ener_balance}.

It is clear from the definitions in \eref{eqn:define_density},
\eref{eqn:define_mom_density}, \eref{eqn:ener_kin} and \eref{eqn:ener_pot} that
the integrals in these equations converge only under appropriate decay
conditions\cite{admal2010} on
$W$. Under these condition, any continuously differentiable vector or 
tensor-valued function defined on the phase space for all $t$ (and satisfying
certain additional decay conditions described in Ref\cite{admal2010}),
we have\footnote{If $\bm{G}$
is a second-order tensor or higher, then the dot product indicates tensor
operating on a vector. Note that in \eref{eqn:reg}, in the interest of brevity,
we are breaking our notation of denoting a second-order tensor operating on a
vector by juxtaposition.}
\begin{subequations}
\label{eqn:reg}
\begin{align}
\int_{\real{3}}\bm{G} \cdot \nabla_{\bm{x}_\alpha}W \, d\bm{x}_\alpha &= -\int_{\real{3}} W \divr_{\bm{x}_\alpha} \bm{G} \, d\bm{x}_\alpha,\label{eqn:reg_1} \\ 
\int_{\real{3}}\bm{G} \cdot \nabla_{\bm{v}_\alpha}W \, d\bm{v}_\alpha &= -\int_{\real{3}} W \divr_{\bm{v}_\alpha} \bm{G} \, d\bm{v}_\alpha. \label{eqn:reg_2}
\end{align}
\end{subequations}
The above identities are repeatedly used in deriving the equation of continuity,
the equation of motion, and the energy balance equation in the
Irving--Kirkwood procedure.
\subsection{General interatomic potentials}
In this section, we describe some properties of interatomic potentials, which
play a crucial role in extending the original Irving--Kirkwood procedure to
multi-body potentials. In addition, it gives new new insights into the original
procedure which was limited to pair potentials. This section is largely based on
Ref\cite{admal2010}, and is briefly summarized here for completeness and to define the necessary notation and terminology.

In general, the internal part of the potential energy, also called the
\emph{interatomic potential energy},
depends on the positions of all particles in the system:
\begin{equation}
\label{eqn:pot_pos_coord}
\pot_{\rm{int}} = \widehat{\pot}_{\rm int}(\bm{x}_1,\bm{x}_2,\dots,\bm{x}_N),
\end{equation}
where the ``hat'' indicates that the functional dependence is on absolute
particle positions (as opposed to distances later on). We assume that
$\widehat{\pot}_{\rm int}:\mathbb{R}^{3N} \to \mathbb{R}$ is a continuously
differentiable function.\footnote{Note that this assumption may fail in systems
undergoing first-order magnetic or electronic phase transformations.} Due to the
invariance of the potential energy with respect to rigid-body motions and
reflections, it can be shown that $\pot_{\rm int}$ in \eref{eqn:pot_pos_coord}
can be expressed as a new function\cite{tadmor2011}
\begin{align}
\pot_{\rm int} = \breve{\pot}_{\rm int}(\cdot),
\end{align}
where the argument of $\breve{\pot}_{\rm int}$ is an $N(N-1)/2$ tuple of
``physically-realizable distances''. Before we describe what this means, we note
that the $N(N-1)/2$ distances between the $N$ particles embedded in $\real{3}$
are not independent. This can be easily seen for any collection of $5$ particles
or more. Therefore, the set of all $N(N-1)/2$ tuples of physically realizable
distances is a proper subset of $\mathbb R^{N(N-1)/2}$. In fact, it is a $(3N-6)$-dimensional manifold called the \emph{shape space} of the system which is defined as
\begin{align}
\mcal{S} := \{(&r_{12}, r_{13}, \dots, r_{1N}, r_{23}, \dots, r_{(N-1)N}) \mid \notag \\
&r_{\alpha \beta} = \vnorm{\bm{x}_\alpha - \bm{x}_\beta}, (\bm{x}_1,\dots,\bm{x}_N) \in \mathbb{R}^{3N}\}.
\end{align}
For example, for a chain of 3 particles in one dimension, with positions 
$x_1<x_2<x_3$, the three distances $(r_{12},r_{13},r_{23})$ must satisfy
$r_{12}+r_{23}=r_{13}$. Values of $(r_{12},r_{13},r_{23})$ that do not
satisfy this constraint are not ``physically realizable'' and are therefore
outside the shape space manifold.

From the above discussion it is clear that the potential energy is
only defined on the shape space of the system. We will soon see that in order to
derive the stress tensor, we need to evaluate partial derivatives like the following:
\begin{align}
\label{eqn:diff_extension}
&\frac{\partial \pot_{\rm int}}{\partial r_{12}} = \notag \\
&\lim_{\epsilon \to 0} 
\frac{\pot_{\rm int}(r_{12}+\epsilon,\dots,r_{N(N-1)}) -\pot_{\rm
int}(r_{12},\dots,r_{N(N-1)})}{\epsilon}.
\end{align}
It is clear that this relation
requires us to evaluate the potential energy outside
the shape space since if $(r_{12},\dots,r_{N(N-1)})$ is on $\mcal{S}$
then by adding $\epsilon$ to one of the distances, we move off it.
Thus, the expression in \eref{eqn:diff_extension} makes
sense only when we extend the function to the neighborhood of the shape
space manifold (see Section 3.4 in Ref\cite{admal2010} for a more detailed
discussion).

This is the reason we now restrict our discussion to those systems for which
there exists a continuously differentiable extension of $\breve{\pot}_{\rm
int}$, defined on the shape space, to $\mathbb{R}^{N(N-1)/2}$. This is
a reasonable assumption because all interatomic potentials used in
practice, for a system of $N$ particles, are either continuously differentiable
functions on $\mathbb{R}^{N(N-1)/2}$, or can easily be extended to one. For
example, the pair potential and the embedded-atom method (EAM) potential
\cite{eam} are continuously differentiable functions on $\mathbb{R}^{N(N-1)/2}$,
while the Stillinger-Weber \cite{stillinger1985} and the Tersoff
\cite{tersoff1988} potentials which depend on the angles between relative
position vectors, can be easily extended to
$\mathbb{R}^{N(N-1)/2}$ by expressing these angles as a
function of distances between particles. Therefore, we assume that there
exists a continuously differentiable function $\pot_{\rm int}:\mathbb{R}^{N(N-1)/2} \to \mathbb{R}$, such that the restriction of $\pot_{\rm int}$ to $\mcal{S}$ is equal to $\breve{\pot}_{\rm int}$:
\begin{align}
\pot_{\rm int} (\bm{s}) = \breve{\pot}_{\rm int}(\bm{s}) \quad
\forall \bm{s}=(r_{12},\dots,r_{(N-1)N}) \in \mcal{S}.
\label{eqn:restriction}
\end{align}                  

An immediate question that arises is whether this extension is unique in a
neighborhood of $\bm{s} \in \mcal{S}$. Note that for $2 \le N \le 4$, $3N-6 =
N(N-1)/2$. Therefore, for $2 \le N \le 4$, for every point $\bm{s} \in \mcal{S}$, there exists a neighborhood in $\mathbb{R}^{N(N-1)/2}$ which lies in $\mcal{S}$. However, for $N>4$, there may be multiple extensions of $\breve{\pot}_{\rm{int}}$.

We will soon see that the quantity evaluated in \eref{eqn:diff_extension}
may differ for different extensions. On the other hand, the \emph{internal} force on any
particle $\alpha$, 
\begin{align}
\label{eqn:def_fi_int}
\bm f_\alpha^{\rm int} &:= -\nabla_{\bm{x}_\alpha} \pot_{\rm int} \notag \\
&=-\nabla_{\bm{x}_\alpha} \widehat{\pot}_{\rm int} 
\end{align}
is uniquely defined for any extension. We next address the possibility of having multiple extensions for the potential
energy by studying the various constraints that the distances between particles
must satisfy in order to be embeddable in $\real{3}$.  We demonstrate,
through a simple example, how multiple extensions for the potential energy
lead to a non-unique decomposition of the force on a particle, which in turn leads to a non-unique pointwise stress tensor.

\subsubsection*{Central-force decomposition and the possibility of alternate extensions}
\label{page:altext}
We first show that the force on a particle can always be decomposed as a sum
of central forces regardless of the nature of the interatomic potential. 
The force on a particle due to internal interactions is
defined in \eref{eqn:def_fi_int}. This can also be evaluated using the
continuously differentiable extension $\pot_{\rm int}$ and the chain rule as 
\begin{align}
\bm{f}^{\rm int}_\alpha (r_{12},\dots,r_{(N-1)N}) &= -\nabla_{\bm{x}_\alpha}
\pot_{\rm{int}}(r_{12},\dots,r_{(N-1)N}) \notag \\
&= \sum_{\substack{\beta \\ \beta \ne \alpha}} \bm{f}_{\alpha\beta}, \label{eqn:f_decomp_general}
\end{align}
where
\begin{equation}
\label{eqn:define_fij}
\bm{f}_{\alpha \beta} := \left \{
\begin{array}{ll}
\frac{\partial\pot_{\rm int}}{\partial r_{\alpha\beta}} \frac{\bm{x}_\beta - \bm{x}_\alpha}{r_{\alpha \beta}} & 
                \mbox{if $\alpha<\beta$}, \\
\frac{\partial\pot_{\rm int}}{\partial r_{\beta\alpha}} \frac{\bm{x}_\beta - \bm{x}_\alpha}{r_{\alpha \beta}} & 
                \mbox{if $\alpha>\beta$},
\end{array}
\right.
\end{equation}
is the contribution to the force on particle $\alpha$ due to the presence of particle $\beta$.

Note that $\bm{f}_{\alpha\beta}$ is parallel to the direction $\bm{x}_\beta -
\bm{x}_\alpha$ and satisfies $\bm{f}_{\alpha \beta}=-\bm{f}_{\beta \alpha}$.
This leads us to the important result that the \emph{internal force on a
particle, for any interatomic potential that has a continuously differentiable
extension, can always be decomposed as a sum of central forces, i.e., forces
parallel to directions connecting the particle to its neighbors}.
This may seem strange to some readers due to the common
confusion in the literature of using the term ``central-force model'' to refer
exclusively to simple pair potentials. In fact, we see that due to the invariance
requirement stated above, \emph{all} interatomic potentials (including
those with explicit bond angle dependence) that can be expressed as a
continuously differentiable function of distance coordinates, are central-force
models. By this we mean that the force on any particle (say $\alpha$) can be
decomposed as a sum (over $\beta$) of terms, $\bm{f}_{\alpha\beta}$, aligned with the vectors
joining particle $\alpha$ with its neighbors and satisfying action and reaction.
The difference between a pair potential and a many-body potential is that in the
former $\bm f_{\alpha\beta}$ only depends on $r_{\alpha\beta}$ whereas in the
latter $\bm f_{\alpha\beta}$ can depend on the distances between all particles.

The next question is how different potential energy extensions affect the
force decomposition in \eref{eqn:f_decomp_general}. We have already seen through
\eref{eqn:def_fi_int} that the force $\bm{f}_\alpha^{\rm int}$ is independent of the particular extension used. However, we show below that the individual terms in the decomposition, $\bm{f}_{\alpha\beta}$, are \emph{not} unique. These terms depend on the manner in which the potential energy, defined on the shape space, is extended to its neighborhood in $\mathbb{R}^{N(N-1)/2}$.

In order to construct different extensions, we use the geometric constraints
that the distances have to satisfy in order for them to be embeddable in
$\real{3}$.\footnote{We thank Ryan Elliott for suggesting this line of
thinking.} The nature of these constraints is studied in the field of
\emph{distance geometry}, which describes the geometry of sets of points in
terms of the distances between them. One of the main results of this theory, is
that the constraints are given by \emph{Cayley-Menger determinants}, which are
related to the volume of a simplex formed by $N$ points in an $N-1$ dimensional
space. The Cayley--Menger determinant corresponding to $N$ particles is given by
\begin{align}
\chi(\zeta_{12},\dots,&\zeta_{1N},\zeta_{23},\dots,\zeta_{(N-1)N})  \notag \\
&=\det 
\begin{bmatrix}
0             & s_{12} & s_{13} & \cdots & s_{1N} &      1 \\
s_{12} &             0 & s_{23} & \cdots & s_{2N} &      1 \\
s_{13} & s_{23} &             0 & \cdots & s_{3N} &      1 \\
       \vdots &        \vdots &        \vdots &        &        \vdots & \vdots \\
s_{1N} & s_{2N} & s_{3N} & \cdots &             0 &      1 \\
            1 &             1 &             1 & \cdots &             1 &      0 
\end{bmatrix},
\label{eqn:cayley}
\end{align}
where $s_{\alpha \beta} = \zeta_{\alpha \beta}^2$.

In the following example we restrict ourselves to one dimension since the
resulting expressions are short and easy to manipulate, although this example
can be readily extended to any dimension. It is easy to see that in one dimension the
number of independent coordinates are $N-1$ and for $N>2$ the number of
interatomic distances exceeds the number of independent coordinates. Therefore,
for simplicity, consider as before a system consisting of three particles interacting
in one dimension. The standard pair potential representation for this system,
which is defined for all $\zeta_{12}$, $\zeta_{13}$ and $\zeta_{23}$ is given by
\begin{equation}
\pot_{\rm{int}}(\zeta_{12},\zeta_{13},\zeta_{23}) = 
\phi_{12}(\zeta_{12}) + \phi_{13}(\zeta_{13}) + \phi_{23}(\zeta_{23}).
\label{eqn:pair_pot}
\end{equation}                                                           
We noted earlier that the distances between particles are geometrically constrained by the requirement
that one of the distance is equal to the sum of the other two.
In spite of this constraint, $\pot_{\rm int}$ is defined
for all values of $(\zeta_{12},\zeta_{13},\zeta_{23})$.
This clearly shows that the pair potential is already an extension.
Since the calculation gets unwieldy, let us again consider the special case where the
particles are arranged to satisfy $x_1<x_2<x_3$, for which
$r_{13}=r_{12}+r_{23}$. Using \eref{eqn:f_decomp_general}, the internal force,
$f_1^{\rm{int}}$, evaluated at this configuration, is decomposed as
\begin{align}
f_1^{\rm{int}}(r_{12},r_{13},r_{23}) = -\frac{d\pot_{\rm{int}}}{dx_1} &= -\frac{d\phi_{12}}{dx_1}-\frac{d\phi_{13}}{dx_1}  \notag \\
&= \phi'_{12}(r_{12})  + \phi'_{13}(r_{13}) \notag \\
& =: f_{12} + f_{13}. \label{eqn:define_fij_pair}
\end{align}                         
We now construct an alternate extension to the standard pair potential
representation given in \eref{eqn:pair_pot}. This is done through the
Cayley-Menger determinant corresponding to a cluster of three points, which
follows from \eref{eqn:cayley} as
\begin{align}
\chi(\zeta_{12},\zeta_{13},\zeta_{23}) &= (\zeta_{12}-\zeta_{13}-\zeta_{23})(\zeta_{23}-\zeta_{12}-\zeta_{13}) \notag \\
& \quad \times (\zeta_{13}-\zeta_{23}-\zeta_{12})(\zeta_{12}+\zeta_{13}+\zeta_{23}). \notag
\end{align}
Since the Cayley--Menger determinant is related to the area formed by the three
particles, and the three particles are restricted to be in one-dimension, it
follows that
\begin{align}
\chi(r_{12},r_{13},r_{23}) = 0.
\label{eqn:cayley_1d}
\end{align}
Using the identity in \eref{eqn:cayley_1d}, an alternate extension $\pot^{\mcal{A}}_{\rm{int}}$ is constructed:
\begin{align}
\pot^{\mcal{A}}_{\rm{int}}(\zeta_{12},\zeta_{13},\zeta_{23}) &= \pot_{\rm int}(\zeta_{12},\zeta_{13},\zeta_{23}) + \chi(\zeta_{12},\zeta_{13},\zeta_{23}).
\label{eqn:pot_rep_2}
\end{align}
Note that $\pot^{\mcal{A}}_{\rm{int}}$ is indeed an extension because from \eref{eqn:cayley_1d} it is clear that $\pot^{\mcal{A}}_{\rm{int}}$ is equal to $\pot_{\rm{int}}$ at every point on the shape space of the system and it is continuously differentiable because $\chi(\zeta_{12},\zeta_{13},\zeta_{23})$, being a polynomial, is infinitely differentiable. Let us now see how the internal force, $f_1^{\rm int}$, for the special configuration considered in this example, is decomposed using the new extension:
\begin{widetext}
\begin{align}
f_1^{\rm{int}} = -\frac{d\pot_{\rm{int}}^{\mcal{A}}}{dx_1} &= 
-\frac{d\pot_{\rm{int}}}{dx_1}  - \frac{d\chi}{dx_1} \notag \\
&= \left (\phi'_{12} - \frac{\partial \chi}{\partial \zeta_{12}}(\bm{s}) \frac{\partial \zeta_{12}}{\partial x_1}(\bm{s}) \right ) +
\left( \phi'_{13} - \frac{\partial \chi}{\partial \zeta_{13}}(\bm{s}) \frac{\partial \zeta_{13}}{\partial x_1}(\bm{s}) \right ) \notag \\
&= \left (f_{12}-8r_{12}r_{23}(r_{12} + r_{23}) \right ) + \left ( f_{13}+8r_{12}r_{23}(r_{12} + r_{23}) \right ) \notag \\
&=: \tilde{f}_{12} + \tilde{f}_{13}, \label{eqn:f1_decomp}
\end{align}
\end{widetext}
where in the above equation $\bm s=(r_{12},r_{13},r_{23})$ is a 
point in the shape space $\mathcal{S}$. 
It is clear from \eref{eqn:define_fij_pair} and \eref{eqn:f1_decomp} that the central-force decomposition is not the same for the two representations, i.e., $f_{12} \ne \tilde{f}_{12}$ and $f_{13} \ne \tilde{f}_{13}$, however the force on particle 1, $f_1^{\rm int}$, is the same in both cases as expected.

\subsection{Equation of Motion and the stress tensor}
\label{sec:s_motion}
The equation of motion and the stress tensor for multi-body potentials has been
extensively studied in the authors' previous work\cite{admal2010}. We now
present those parts of the derivation which are necessary to derive the energy
equation in \sref{sec:ener_balance}. The equation of motion from continuum mechanics is given by \cite{malvern}
\begin{eqnarray}
\label{eqn:motion}
\frac{\partial(\rho \bm{v})}{\partial t} + \divr_{\bm{x}}(\rho \bm{v} \otimes \bm{v}) = \divr_{\bm{x}}\bm{\sigma} + \bm{b},
\end{eqnarray}
where $\bm{\sigma}$ is the Cauchy stress tensor and $\bm{b}$ is the body force field.
Using Liouville's equation and substituting in the definitions for mass density and the
velocity fields defined in \eref{eqn:define_density} and
\eref{eqn:define_velocity}, respectively, into \eref{eqn:motion}, it can be shown that the stress tensor
and the body force field must satisfy 
\begin{align}
\divr_{\bm{x}}\bm{\sigma} + \bm{b} = &-\sum_{\alpha} m_\alpha \divr_{\bm{x}}
\left \langle (\bm{v}_\alpha^{\rm{rel}} \otimes \bm{v}_\alpha ^{\rm{rel}}) W
\mid \bm{x}_\alpha = \bm{x} \right \rangle \notag \\
&- \sum_{\alpha} \left \langle W \nabla_{\bm{x}_\alpha} \pot_{\rm int} \mid \bm{x}_\alpha
= \bm{x} \right \rangle \notag \\
&- \sum_{\alpha} \left \langle W \nabla_{\bm{x}_\alpha} \pot_{\rm ext} \mid \bm{x}_\alpha
= \bm{x} \right \rangle,
\label{eqn:motion_5}
\end{align}
where
\begin{align}
\bm v_\alpha^{\rm rel} := \bm v_\alpha - \bm v
\label{eqn:v_alpha}
\end{align}
is the velocity of particle $\alpha$ relative to the pointwise velocity field.
It is natural to associate $\pot_{\rm{ext}}$ with the body force field $\bm{b}$ in \eref{eqn:motion_5}. We therefore define $\bm{b}(\bm{x},t)$ as
\begin{equation}
\label{eqn:body}
\bm{b}(\bm{x},t) := - \sum_{\alpha} \left \langle W \nabla_{\bm{x}_\alpha} \pot_{\rm{ext}} \mid \bm{x}_\alpha = \bm{x} \right \rangle.
\end{equation}
Substituting \eref{eqn:body} into \eref{eqn:motion_5}, we have
\begin{align}
\divr_{\bm{x}} \bm{\sigma} = &-\sum_{\alpha} m_\alpha \divr_{\bm{x}} \left
\langle (\bm{v}_\alpha ^{\rm{rel}} \otimes \bm{v}_\alpha^{\rm{rel}}) W \mid
\bm{x}_\alpha = \bm{x} \right \rangle \notag \\
&- \sum_{\alpha} \left \langle W \nabla_{\bm{x}_\alpha} \pot_{\rm{int}}  \mid \bm{x}_\alpha = \bm{x} \right \rangle.
\label{eqn:motion_6}
\end{align}
From \eref{eqn:motion_6}, we see that the pointwise stress tensor has two contributions:
\begin{equation}
\label{eqn:stress_split}
\bm{\sigma}(\bm{x},t) = \bm{\sigma}_{\rm{k}}(\bm{x},t) + \bm{\sigma}_{\rm{v}}(\bm{x},t),
\end{equation}
where $\bm{\sigma}_{\rm{k}}$ and $\bm{\sigma}_{\rm{v}}$ are, respectively, the \emph{kinetic} and \emph{potential} parts of the pointwise stress.  The kinetic part is given by
\begin{equation}
\bm{\sigma}_{\rm{k}}(\bm{x},t) = -\sum_{\alpha} m_\alpha \left \langle (\bm{v}_\alpha^{\rm{rel}} \otimes \bm{v}_\alpha ^{\rm{rel}}) W \mid \bm{x}_\alpha = \bm{x} \right \rangle.
\label{eqn:stress_kinetic}
\end{equation}
It is evident that the kinetic part of the stress tensor is symmetric. The
kinetic stress reflects the momentum flux associated with the vibrational
kinetic energy portion of the internal energy.

Continuing with \eref{eqn:motion_6}, the potential part of the stress must satisfy the following differential equation:
\begin{equation}
\divr_{\bm{x}} \bm{\sigma}_{\rm{v}}(\bm{x},t) = \sum_{\alpha} \left \langle W \bm{f}^{\rm int}_\alpha \mid \bm{x}_\alpha = \bm{x} \right \rangle,
\label{eqn:stress_force_differential}
\end{equation}
Equation~\eref{eqn:stress_force_differential} needs to be solved in order to
obtain an explicit form for $\bm{\sigma}_{\rm{v}}$. In the original paper of
Irving and Kirkwood \cite{ik1950}, a solution to
\eref{eqn:stress_force_differential} was obtained for the special case of pair
potential interactions by applying a Taylor expansion to the Dirac delta
distribution. In contrast, Noll
showed that a closed-form solution for $\bm{\sigma}_{\rm{v}}$ can be obtained by
recasting the right-hand side in a different form and applying a lemma proved in
 Ref~\cite{noll1955}.  We proceed with Noll's approach, except we place no
restriction on the nature of the interatomic potential energy $\pot_{\rm{int}}$.

\subsubsection*{Derivation of the pointwise stress tensor}
We substitute the force
decomposition given in \eref{eqn:f_decomp_general} corresponding to a
continuously differentiable extension into
the potential part of the pointwise stress tensor in
\eref{eqn:stress_force_differential} to obtain
\begin{equation}
\label{eqn:stress_force_differential_*}
\divr_{\bm{x}} \stress_{\rm{v}}(\bm{x},t) = \sum_{\substack{\alpha,\beta \\ \alpha \ne \beta}} \langle W \bm{f}_{\alpha \beta} \mid \bm{x}_\alpha = \bm{x} \rangle.
\end{equation}
On using the identity 
\begin{equation}
\left \langle \bm{f}_{\alpha\beta} W \mid \bm{x}_\alpha = \bm{x} \right \rangle = \int_{\real{3}} \left \langle \bm{f}_{\alpha\beta} W \mid \bm{x}_\alpha = \bm{x}, \bm{x}_\beta = \bm{y} \right \rangle \, d \bm{y},
\end{equation}
equation \eref{eqn:stress_force_differential_*} takes the form
\begin{equation}
\label{eqn:stress_force_differential_**}
\divr_{\bm{x}} \bm{\sigma}_{\rm{v}}(\bm{x},t) = \int_{\real{3}} \sum_{\substack{\alpha,\beta \\ \alpha \ne \beta}} \left \langle W \bm{f}_{\alpha\beta} \mid \bm{x}_\alpha = \bm{x}, \bm{x}_\beta = \bm{y} \right \rangle \, d\bm{y}.
\end{equation}
We now note the following lemma due to Noll, which will be used to obtain a closed-form
solution to the output fields derived in the Irving--Kirkwood procedure.
\begin{lemma}
Let $\bm{f}(\bm{v},\bm{w})$ be a tensor-valued function of two vectors $\bm{v}$ and $\bm{w}$, which satisfies the following three conditions:
\begin{enumerate}
\label{lem1}
\item $\bm{f}(\bm{v},\bm{w})$ is defined for all $\bm{v}$ and $\bm{w}$ and is continuously differentiable \label{condition_1}.
\item There exists a $\delta > 0$, such that the auxiliary function $\bm{g}(\bm{v},\bm{w})$, defined through
\begin{equation}
\bm{g}(\bm{v},\bm{w}):=\bm{f}(\bm{v},\bm{w}) \vnorm{\bm{v}} ^{3+\delta} \vnorm{w}^{3+\delta},
\end{equation}
and its gradients $\nabla_{\bm{v}} \bm{g}$ and $\nabla_{\bm{w}}\bm{g}$ are bounded.\label{condition_2}
\item $\bm{f}(\bm{v},\bm{w})$ is antisymmetric, i.e.,
\begin{equation}
\bm{f}(\bm{v},\bm{w}) = -\bm{f}(\bm{w},\bm{v}).
\label{eqn:anti_sym_1}
\end{equation}
\label{condition_3}
\end{enumerate}
Under the above conditions, the following equation holds:\footnote{The expression in Noll's paper appears transposed relative to \eref{eqn:lemma1}. This is because the gradient and divergence operations used by Noll are the transpose of our definitions.}
\begin{multline}
\label{eqn:lemma1}
\int_{\bm{y} \in \real{3}} \bm{f}(\bm{x},\bm{y}) \,d\bm{y} =
\\
-\frac{1}{2} \divr_{\bm{x}} \int_{\bm{z} \in \real{3}} 
\left [ \int_{s=0}^{1} \bm{f}(\bm{x} + s\bm{z},\bm{x}-(1-s)\bm{z}) \, ds \right ]\otimes \bm{z} \, d\bm{z}.
\end{multline}
\end{lemma} 
It is clear that being anti-symmetric, the integrand in the right-hand side of
\eref{eqn:stress_force_differential_**} satisfies all the necessary conditions
for the application of Lemma \ref{lem1}. Conditions (1) and (2) are satisfied
through the regularity conditions on $W$. Therefore, using Lemma \ref{lem1}, we have
\begin{widetext}
\begin{align}
\label{eqn:stress_force_general}
\bm{\sigma}_{\rm{v}}(\bm{x},t) &= \frac{1}{2} \sum_{\substack{\alpha,\beta \\
\alpha \neq \beta}} \int_{\real{3}} \int_{s=0}^{1} \left \langle
-\bm{f}_{\alpha\beta} W \mid \bm{x}_\alpha=\bm{x}+s\bm{z},\bm{x}_\beta = \bm{x}
- (1-s)\bm{z} \right \rangle \, ds \otimes \bm{z} \, d\bm{z}. \notag \\
\end{align}
\end{widetext}
\begin{figure}
\centering
\includegraphics[scale=0.4]{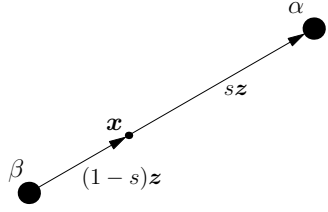}
\caption{A schematic diagram helping to explain the vectors appearing in the pointwise potential stress expression in \eref{eqn:stress_force_general}. The bond $\alpha$--$\beta$ is defined by the vector $\bm{z}$. When $s=0$, atom $\alpha$ is located at point $\bm{x}$, and when $s=1$, atom $\beta$ is located at $\bm{x}$.}
\label{fig:stressbond}
\end{figure} 

\medskip
The expression for the potential part of the pointwise stress tensor in \eref{eqn:stress_force_general} is a general result applicable to all interatomic potentials. We make some important observations regarding this expressions below:
\begin{enumerate}

\item The expression for $\bm{\sigma}_{\rm{v}}$ given in
\eref{eqn:stress_force_general} has an easy
interpretation. $\bm{\sigma}_{\rm{v}}$ at a point $\bm{x}$ is the superposition
of the expectation values of the forces in all possible bonds passing through
$\bm{x}$. The variable $\bm{z}$ selects a bond length and direction and the
variable $s$ slides the bond through $\bm{x}$ from end to end (see \fref{fig:stressbond}).

\item $\bm{\sigma}_{\rm{v}}$ is symmetric. This is clear because 
$\bm{f}_{\alpha\beta}$ is parallel to $\bm z$ and $\bm{z}\otimes\bm{z}$ is symmetric.
Since the kinetic part of the stress in \eref{eqn:stress_kinetic} is also
symmetric, we can conclude that the \emph{pointwise stress tensor is symmetric
for all interatomic potentials}.

\item Since $\bm{\sigma}_{\rm{v}}$ depends on the nature of the force decomposition and different extensions of a given potential energy can result in different force decompositions, we conclude that the pointwise stress tensor is \emph{non-unique} for all interatomic potentials (including the pair potential).

\end{enumerate}
The non-uniqueness of the pointwise stress tensor also plays an important
role in the energy equation, derived in the next section, since the stress appears in it.

\subsection{Equation of energy balance}
\label{sec:ener_balance}
The energy balance equation from continuum mechanics is given by
\begin{align}
\label{eqn:energy_bal}
\frac{\partial \rho \epsilon}{\partial t} + \divr_{\bm x}(\bm q - \bm \sigma
\bm v + \rho \epsilon \bm v) = \bm 0,
\end{align}
where $\epsilon$ is the specific internal energy and $\bm{q}$ is the heat
flux vector.  We saw in the previous
section that the Irving--Kirkwood procedure extended to general
interatomic potentials yields various possible definitions for the stress
tensor. In a similar vein, we hope to use this extended procedure to derive possible
definitions for the heat flux vector for arbitrary multi-body potentials. Before
that, let us look at the definition for the heat flux vector given by the
original Irving--Kirkwood procedure for the case of a pair potential. The heat flux
vector in this case is decomposed as 
\begin{align}
\label{eqn:heat_flux}
\bm q := \bm q_{\rm k} + \bm q_{\rm T} + \bm q_{\rm v},
\end{align}
where 
\begin{align}
\bm q_{\rm k} &:= \frac{1}{2}\sum_\alpha m_\alpha \langle \vnorm{\bm v_\alpha^{\rm
rel}}^2 \bm v_\alpha^{\rm rel} W \mid \bm x_\alpha = \bm x \rangle,
\label{eqn:qk} \\
\bm q_{\rm T} &:= \frac{1}{2}\sum_{\alpha}
\langle \bm v_\alpha^{\rm rel} \pot_\alpha W \mid \bm x_\alpha = \bm x
\rangle, \label{eqn:qt}
\end{align}
and
\begin{widetext}
\begin{align}
\bm q_{\rm v} := -\frac{1}{2} \sum_{\substack{\alpha, \beta \\ \alpha \ne \beta}}
\int_{\bm z \in \real{3}} \frac{\bm z}{\vnorm{\bm z}} \bm z \cdot
\phi'_{\alpha\beta} \int_{s=0}^{1}
\left \langle  \left (\frac{\bm v_\alpha +
\bm v_\beta}{2}-\bm v \right ) W \mid \bm x_\alpha = \bm x+s\bm z, \bm x_\beta = \bm x
- (1-s)\bm z \right \rangle \, ds \, d\bm z, \label{eqn:qv}
\end{align}
\end{widetext}
represent the kinetic part, transport part, and the potential part of the heat
flux vector respectively. It was shown by Noll \cite{noll1955} that if the heat
flux vector is defined according to \eref{eqn:heat_flux},
then along with the definition for the specific internal energy given in \eref{eqn:ener},
and Lemma \ref{lem1}, the energy balance equation \eref{eqn:energy_bal} is
identically satisfied. We can now try to extend this procedure to arbitrary
multi-body potentials by defining a potential energy extension and repeat the
steps given in Ref\cite{noll1955}. But before we can do so, we must grapple with the
ambiguity that arises in the definition for the potential part of the specific
energy, $\epsilon_{\rm v}$, given in \eref{eqn:ener_pot}. As mentioned at the end of
\sref{sec:phase}, in order to define $\epsilon_{\rm v}$ for multi-body
potentials we must give a precise definition for the energy of each particle
$\pot_\alpha$ as done in \eref{eqn:def_pot_alpha} for a pair potential. Even
for the case of identical particles, it is not \emph{a priori}
clear how to distribute the energy between the particles for a multi-body
potential. It is clear from
\eref{eqn:qt} that this results in an ambiguous definition for
$\bm q_{\rm T}$ which depends on the definition for $\pot_\alpha$. Moreover, one
would expect that the definitions for the pointwise fields should be invariant
with respect to addition of any constant to the potential energy. It is clear
that all the definitions discussed so far satisfy this invariance except for
\eref{eqn:qt}. Therefore a question that naturally arises is, whether the decomposition
of energy is necessary to derive the energy balance equation. An alternate approach
which we think is more reasonable comes from a paper by
Murdoch\cite{murdoch1994} in his spatial averaging procedure. Here we adapt this
approach to the Irving--Kirkwood procedure.

\subsubsection*{An alternate derivation of the energy balance equation}
The alternate derivation for the energy balance equation in this section leads
to an expression for the heat flux vector which
does not contain the transport part. Moreover, this derivation applies
to any multi-body potential with a continuously differentiable extension. Under
this alternate derivation we have the following input and output fields:
\begin{align}
\label{eqn:fields_new}
\textbf{Input}  \quad & \quad \qquad \textbf{Output} \notag \\
\left \{ \begin{array}{l} 
\rho \\
\bm v \\
\epsilon_{\rm k} \end{array} \right \} 
&\rightarrow
\left \{ \begin{array}{c} 
\epsilon_{\rm v} \\
\bm b \\
\stress \\
\bm q = \bm q_{\rm k} + \bm q_{\rm v} \end{array} \right \}
\end{align}
We consider the terms in \eref{eqn:energy_bal}, beginning 
with $\rho \epsilon_{\rm k}$ defined in \eref{eqn:ener_kin}.
For simplicity, we assume $\pot_{\rm ext}=0$. We have
\begin{align}
\label{eqn:ener_kin_partial}
\frac{\partial E_{\rm k}}{\partial t} = \frac{1}{2}\sum_\alpha m_\alpha
\left \langle \vnorm{\bm v_\alpha}^2 \frac{\partial W}{\partial t} \mid \bm x_\alpha =
\bm x \right \rangle,
\end{align}
where $E_{\rm k} := \rho \epsilon_{\rm k}$. Using Liouville's equation given in
\eref{eqn:liouville}, we obtain
\begin{widetext}
\begin{align}
\frac{\partial E_{\rm k}}{\partial t} &= \frac{1}{2} \sum_\alpha m_\alpha
\left \langle \vnorm{\bm v_\alpha}^2 \sum_\beta \left (-\bm v_\beta \cdot \nabla_{\bm
x_\beta} W + \frac{\nabla_{\bm x_\beta} \pot}{m_\beta} \cdot \nabla_{\bm v_\beta} W \right )\mid \bm
x_\alpha = \bm x \right \rangle \notag \\
&= -\frac{1}{2} \sum_\alpha m_\alpha \left \langle \vnorm{\bm v_\alpha}^2 \bm v_\alpha
\cdot \nabla_{\bm x_\alpha} W \mid \bm x_\alpha = \bm x \right \rangle -
\frac{1}{2} \sum_\alpha \left \langle \vnorm{\bm v_\alpha}^2 \bm f_\alpha^{\rm
int} \cdot
\nabla_{\bm v_\alpha} W \mid \bm x_\alpha = \bm x \right \rangle,\notag \\
&=: \bm q_1 + \bm q_2, \label{eqn:def_q1q2}
\end{align}
\end{widetext}
where we have used the identities \eref{eqn:reg_1} and \eref{eqn:reg_2}. Now note that the term
$\vnorm{\bm v_\alpha}^2 \bm v_\alpha$ can be written as
\begin{align}
\label{eqn:valpha_id}
\vnorm{\bm v_\alpha}^2 \bm v_\alpha = \vnorm{\bm v_\alpha^{\rm rel}}^2 \bm v_\alpha^{\rm rel}
&+ 2 (\bm v_\alpha^{\rm rel} \otimes \bm v_\alpha^{\rm rel} ) \bm v + \notag \\
&\bm v \vnorm{\bm v_\alpha}^2 + \vnorm{\bm v}^2 \bm v_\alpha^{\rm rel}.
\end{align}
Consider $\bm q_1$, the first term of \eref{eqn:def_q1q2}. Using
\eref{eqn:valpha_id} and the definitions for $\bm q_{\rm k}$, $\stress_{\rm k}$
and $E_{\rm k}$ given in \eref{eqn:qk}, \eref{eqn:stress_kinetic} and
\eref{eqn:ener_kin} respectively, $\bm q_1$ can be expressed as
\begin{align}
\label{eqn:def_q1}
\bm q_1 &= -\divr_{\bm x} ( \bm q_{\rm k} -\stress_{\rm k} \bm v + E_{\rm
k} \bm v ) - \notag \\
& \qquad \frac{1}{2} \vnorm{\bm v}^2 \divr_{\bm x} \sum_\alpha m_\alpha \left \langle
\bm v_\alpha^{\rm rel} W \mid \bm x_\alpha = \bm x \right \rangle \notag \\
&=-\divr_{\bm x} ( \bm q_{\rm k} -\stress_{\rm k} \bm v + E_{\rm
k} \bm v ),
\end{align}
since $\sum_\alpha m_\alpha \langle \bm v_\alpha^{\rm rel} W \mid \bm x_\alpha =
\bm x \rangle = \bm 0$. Now consider $\bm q_2$, the second term of
\eref{eqn:def_q1q2}. Integrating by parts, and using the regularity conditions on
$W$, $\bm q_2$ takes the form
\begin{align}
\label{eqn:def_q2}
\bm q_2 = \sum_\alpha \langle \bm v_\alpha \cdot \bm f_\alpha^{\rm int} W \mid \bm x_\alpha = \bm
x \rangle.
\end{align}
Using \eref{eqn:def_q1} and \eref{eqn:def_q2}, \eref{eqn:def_q1q2} becomes
\begin{align}
&\frac{\partial E_{\rm k}}{\partial t} \notag \\
&= -\divr_{\bm x} (\bm q_{\rm k} -
\stress_{\rm k} \bm v + E_{\rm k} \bm v) + \sum_\alpha \langle \bm
v_\alpha \cdot \bm f_\alpha^{\rm int} W \mid \bm x_\alpha = \bm x \rangle \notag \\
&= -\divr_{\bm x} (\bm q_{\rm k} - \stress_{\rm k} \bm v + E_{\rm k} \bm
v) + \sum_\alpha \langle \bm v_\alpha^{\rm rel} \cdot \bm f_\alpha^{\rm int} W \mid \bm
x_\alpha = \bm x \rangle \notag \\
& \quad + \left [ \sum_\alpha \langle \bm f_\alpha^{\rm int} W \mid \bm
x_\alpha = \bm x \rangle \right ] \cdot \bm v. \label{eqn:differential_ek1}
\end{align}
We know from the  momentum balance equation (see
\eref{eqn:stress_force_differential}) that
\begin{align}
\label{eqn:stress_force_differential_***}
\divr_{\bm x} \stress_{\rm v} = \sum_\alpha \langle \bm f_\alpha^{\rm int} W \mid \bm
x_\alpha = \bm x \rangle.
\end{align}
Using \eref{eqn:stress_force_differential_***}, together with the identity
\begin{align*}
\divr_{\bm x} (\bm T \bm b) = \divr(\bm T) \cdot \bm b + \bm T : \nabla_{\bm x}
\bm b,
\end{align*}
where $T$ and $\bm b$ are continuously differentiable tensor and vector-valued functions of $\bm x$
respectively, and noting that $\stress = \stress_{\rm k} +
\stress_{\rm v}$, \eref{eqn:differential_ek1} can be rewritten as
\begin{align}
\frac{\partial E_{\rm k}}{\partial t} &= -\divr_{\bm x} (\bm q_{\rm k} -
\stress \bm v + E_{\rm k} \bm v ) \notag \\
&+ \sum_\alpha \langle \bm v_\alpha^{\rm
rel} \cdot \bm f_\alpha^{\rm int} W \mid \bm x_\alpha = \bm x \rangle - \stress :
\nabla_{\bm x} \bm v. \label{eqn:differential_ek2}
\end{align}
Now, consider the middle term on the right-hand side of \eref{eqn:differential_ek2} which
is given by
\begin{align}
\label{eqn:def_q3}
\bm q_3 &:= \sum_{\alpha} \langle \bm
v_{\alpha}^{\rm rel} \cdot \bm f_\alpha^{\rm int} W \mid \bm x_\alpha = \bm x
\rangle.
\end{align}
Substituting the force decomposition given in \eref{eqn:f_decomp_general}
corresponding to a continuously differentiable extension, into
\eref{eqn:def_q3}, we obtain
\begin{align}
\bm q_3 &= \sum_{\substack{\alpha,\beta \\ \alpha \ne \beta}} \langle \bm
v_{\alpha}^{\rm rel} \cdot \bm f_{\alpha\beta} W \mid \bm x_\alpha = \bm x
\rangle \notag \\
& =\sum_{\substack{\alpha,\beta \\ \alpha \ne \beta}} \int_{\real{3}} \langle
\bm v_\alpha^{\rm rel} \cdot \bm f_{\alpha\beta} W \mid \bm x_\alpha = \bm x, \bm
x_\beta = \bm y \rangle \, d\bm y \notag \\
&= \int_{\real 3} \left [ g_{\rm S}(\bm x,\bm y) + g_{\rm {AS}}(\bm x,\bm y)
\right ] \, d\bm y,
\label{eqn:g+g-}
\end{align}
where
\begin{multline}
g_{\rm S}(\bm{x},\bm{y}) = \\
\frac{1}{2} \sum_{\substack{\alpha,\beta \\ \alpha \ne
\beta}} \langle (\bm f_{\alpha\beta} \cdot \bm v_\alpha^{\rm rel} + \bm f_{\beta\alpha}
\cdot \bm v_\beta^{\rm rel} ) W 
\mid \bm x_\alpha = \bm x, \bm x_\beta = \bm y
\rangle  \label{eqn:g+},
\end{multline}
\begin{multline}
g_{\rm {AS}}(\bm{x},\bm{y}) = \\
\frac{1}{2} \sum_{\substack{\alpha,\beta \\ \alpha \ne
\beta}} \langle (\bm f_{\alpha\beta} \cdot \bm v_\alpha^{\rm rel} - \bm
f_{\beta\alpha} \bm v_\beta^{\rm rel}) W
\mid \bm x_\alpha = \bm x, \bm x_\beta
= \bm y \rangle \label{eqn:g-}.
\end{multline}
It is easy to check that $g_{\rm {A}S}(\bm x,\bm y) = -g_{\rm {A}S}(\bm y,\bm x)$, i.e., it is
antisymmetric with respect to its arguments. Moreover, the second integrand on
the right-hand side of \eref{eqn:g+g-} satisfies all the necessary conditions for the
applications of Lemma \ref{lem1}. Using Lemma
\ref{lem1}, we can express the second integral in \eref{eqn:g+g-} as
\begin{align}
\label{eqn:int_g-}
\int_{\real 3} g_{\rm {A}S}(\bm x,\bm y) \, d\bm y = -\divr_{\bm x} \bm q_{\rm v}.
\end{align}
where
\begin{widetext}
\begin{align}
\bm q_{\rm v} := \frac{1}{2} \sum_{\substack{\alpha, \beta \\ \alpha \ne \beta}}
\int_{\bm z \in \real{3}} \bm z \int_{s=0}^{1}
\left \langle \bm f_{\alpha\beta} \cdot \left (\frac{\bm v_\alpha +
\bm v_\beta}{2}-\bm v \right ) W \mid \bm x_\alpha = \bm x+s\bm z, \bm x_\beta = \bm x
- (1-s)\bm z \right \rangle \, ds \, d\bm z. \label{eqn:qv_new}
\end{align}
\end{widetext}
Substituting \eref{eqn:int_g-} into \eref{eqn:g+g-}, and noting that 
\begin{align}
\int_{\real{3}} g_{\rm S}(\bm x,\bm y) \, d\bm y &= \frac{1}{2}
\sum_{\substack{\alpha,\beta \\ \alpha \ne \beta}} \langle \bm f_{\alpha\beta}
\cdot (\bm v_\alpha-\bm v_\beta)W \mid \bm x_\alpha=\bm x \rangle, \notag \\
& =: \bar g_{\rm S}(\bm x,t)
\end{align}
we obtain
\begin{align}
\frac{\partial E_{\rm k}}{\partial t} = -\divr_{\bm x} [\bm q_{\rm k} + \bm
q_{\rm v} - \stress \bm v + E_k \bm v] - \stress : \nabla_{\bm x} \bm v
+ \bar g_{\rm S}(\bm x,t). \label{eqn:differential_ek3}
\end{align}
Now, recall the energy equation of continuum thermodynamics in
\eref{eqn:energy_bal}. 
Subtracting \eref{eqn:differential_ek3} form \eref{eqn:energy_bal}, we obtain
\begin{align}
\label{eqn:differential_ek4}
\frac{\partial (\rho \epsilon_{\rm v})}{\partial t} = -\divr_{\bm x} (\bm q -
\bm q_{\rm k} - \bm q_{\rm v} + \rho
\epsilon_{\rm v} \bm v) +
\stress : \nabla_{\bm x} \bm v - \bar g_{\rm S}(\bm x,t).
\end{align}
The following step is a crucial part of our derivation.  
Note that in contrast to the original Irving--Kirkwood derivation,
the transport part of the heat flux, $\bm q_{\rm T}$, 
does not appear here. \emph{We can therefore identify the heat flux vector $\bm q$ with 
$\bm q_{\rm k}+\bm q_{\rm v}$}, i.e., 
\begin{align}
\bm q:=\bm q_{\rm k} + \bm q_{\rm v}.
\label{eqn:def_q_new}
\end{align}
Thus,
\eref{eqn:differential_ek4} reduces to
\begin{align*}
\frac{\partial (\rho \epsilon_{\rm v})}{\partial t} = -\divr_{\bm x} (\rho
\epsilon_{\rm v} \bm v) +
\stress : \nabla_{\bm x} \bm v - \bar g_{\rm S}(\bm x,t),
\end{align*}
which implies that
\begin{align}
\epsilon_{\rm v} \frac{\partial \rho}{\partial t} + \rho \frac{\partial
\epsilon_{\rm v}}{\partial t} &= \epsilon_{\rm v} \divr_{\bm x}(\rho \bm v) - \rho
\nabla \epsilon_{\rm v} \cdot \bm v \notag \\
&+ \stress:\nabla_{\bm x} \bm v - 
\bar{g}_{\rm S}(\bm x).  \label{eqn:compare}
\end{align}
Using the equation of continuity, \eref{eqn:compare}
simplifies to 
\begin{align}
\rho \left ( \frac{\partial \epsilon_{\rm v}}{\partial t} + \nabla \epsilon_{\bm
v} \cdot \bm v \right ) = \stress:\nabla_{\bm x} \bm v - \bar g_{\rm S}(\bm x,t),
\end{align}
which implies that
\begin{align}
\label{eqn:ener_pot_dot}
\rho \dot{\epsilon}_{\rm v} = \stress : \nabla_{\bm x} \bm v - 
\bar g_{\rm S}(\bm x,t).
\end{align}
It is clear from \eref{eqn:ener_pot_dot} that we now have a new definition for the specific
internal energy (similar to the one obtained by Murdoch\cite{murdoch1994}
in the Murdoch procedure) given by
\begin{align}
\label{eqn:ener_pot_new}
\epsilon_{\rm v}(\bm x, t) = \int_0^t \frac{1}{\rho} \left ( \stress:\nabla_{\bm
x} \bm v - \bar g_{\rm S}(\bm x,t) \right )\, dt + c.
\end{align}
This definition does not require a decomposition of the total energy to
individual atoms, i.e., it is independent of a particular choice for
$\pot_\alpha$, contrary to what is observed in the original Irving--Kirkwood
procedure and its generalization to multi-body potentials found in the
literature (see Ref\cite{chen2006,zhang2004,zim2008,marechal1983,torii2008}). 

In summary, we obtained new
definitions for $\epsilon_{\rm v}$ and $\bm q$, which are quite different from
those obtained in the Irving--Kirkwood procedure. We believe that the
new definitions for $\epsilon_{\rm v}$ and $\bm q_{\rm v}$ given in
\eref{eqn:ener_pot_new} and \eref{eqn:def_q_new} respectively are more physically reasonable
as compared to those given in \eref{eqn:ener_pot} and
\eref{eqn:heat_flux} due to the following features which are not observed in the
Irving--Kirkwood procedure or any of its previous generalizations to multi-body potentials:
\begin{enumerate}
\item The definitions for $\bm q$ and $\epsilon_{\rm v}$ given in
\eref{eqn:def_q_new} and \eref{eqn:ener_pot_new} depend on the derivative of the
potential thus making them invariant with respect to changes in the potential
energy by a constant. This is a rather natural thing to expect.
\item The heat flux vector obtained in the alternate derivation does not have 
transport part.  This suggests that we look for numerical experiments which
yield a non-trivial transport part using the original Irving--Kirkwood
procedure. Most of the numerical experiments found in the literature, 
which study the energy balance equation obtained though the Irving--Kirkwood
procedure, lump the transport part into either the kinetic or potential parts of
of the heat flux vector and do not observe it separately.  Hence, there has been no
extensive numerical study of the role of this term.
If indeed the expression for the transport part of the heat flux vector
found in the Irving--Kirkwood procedure always has a negligible
contribution to the heat flux vector, then its existence can be questioned.
Preliminary numerical simulations we conducted to explore this (which are not
reported here) always yielded a negligible transport part.

\item From \sref{sec:s_motion} and \sref{sec:ener_balance}, it follows that the only ambiguity in the
expressions obtained through this modified derivation is the non-uniqueness of
the pointwise stress tensor, which is directly related to the force
decomposition. It was shown in Ref\cite{admal2010} that this non-uniqueness
vanishes in the thermodynamic limit.
\end{enumerate}

\section{Expression for MD simulation}
\label{sec:md}
In the previous sections we saw that various pointwise fields can be
obtained through the Irving--Kirkwood procedure. As noted in Section II,
the pointwise field are not continuum fields. We identify the continuum fields
with macroscopic observables obtained in a two-step process:
\begin{enumerate}
\item A pointwise field is obtained as a statistical mechanics phase average.
\item A macroscopic field is obtained as a spatial average over the pointwise
field.
\end{enumerate}
We have seen that the pointwise fields obtained in the first step are defined as
phase averages with respect to a probability density function. Typically a
molecular dynamics (MD)
simulation is purely deterministic in nature, meaning that at a given instant in time, we
have a complete microscopic description of the system. Due to this knowledge, the
probability density function introduced in the Irving--Kirkwood procedure
reduces to  
a Dirac delta distribution supported on the point in the phase space
corresponding to the state of the system. If $(\bm x^{\rm MD}(t),\bm v^{\rm
MD}(t))$
denotes the evolution of an MD simulation, then the probability density function
$W^{\rm MD}$ corresponding to an MD simulation is given by
\begin{align}
\label{eqn:w_md}
W^{\rm MD}(\bm x,\bm v;t) = \prod_{\alpha} \delta(\bm x_\alpha - \bm x_\alpha^{\rm MD}(t))
\delta(\bm v_\alpha - \bm v_\alpha^{\rm MD}(t)).
\end{align}
Therefore, in an MD setting, the pointwise fields obtained
in step 1 are localized to the particle positions. 
Next, we spatially average these fields with
respect to a normalized weighting function that has compact support, thus
obtaining expressions for the continuum fields that can be numerically evaluated using the
data generated in a MD simulation. 

\subsection*{Spatial averaging}
A macroscopic quantity is by necessity an average over some spatial region
surrounding the continuum point where it is nominally defined. Thus, if 
$f(\bm{x},t; W)$ is an Irving--Kirkwood- pointwise field, such as density, stress
or internal energy, the corresponding macroscopic field ${f}_w(\bm{x},t)$ 
is given by
\begin{equation}
\label{eqn:define_se_0}
f_w(\bm{x},t) = \int_{\real{3}} w(\bm{y} - \bm{x}) f(\bm{y},t; W) \, d\bm{y},
\end{equation}
where $w(\bm{r})$ is a suitable weighting function.

It is important to note that due to the linearity of the phase averaging in the Irving--Kirkwood procedure, the averaged macroscopic function ${f}_w(\bm{x},t)$ satisfies the same balance equations as does the pointwise measure $f(\bm{x},t)$.

\subsection*{Weighting function} 
The weighting function $w(\bm{r})$ is a
real-valued function\footnote{It was mentioned in Ref\cite{admal2010} that the
weighting function is a positive-valued function based on our interpretation of
it being related to the nature of the experimental probe. We thank
I.~Murdoch for directing us to an alternate interpretation which allows the weighting
function to take on negative values. See Ref\cite{murdoch1994} for details.
} 
with units of $\rm{volume}^{-1}$ 
which satisfies the normalization condition
\begin{equation}
\label{eqn:normal}
\int_{\real{3}} w(\bm{r}) d\bm{r} = 1.
\end{equation}
This condition ensures that the correct macroscopic field is obtained when the
pointwise field is uniform. For a spherically-symmetric function,
$w(\bm{r}) = \hat{w}(r)$, where $r=\vnorm{\bm{r}}$. The normalization condition
in this case is
\[
\int_{0}^{\infty} \hat{w}(r)4 \pi r^2 dr = 1.
\]
The simplest choice for $\hat{w}(r)$ is a spherically-symmetric uniform
function over a specified radius $r_w$, given by
\begin{equation}
\label{eqn:constant_w}
\hat{w}(r) = \left \{ \begin{array}{ll}
1/V_w & \mbox{if $r \leq r_w$},\\
0 & \mbox{otherwise},\end{array} \right. 
\end{equation}
where $V_w = \frac{4}{3}\pi r_{w}^{3}$ is the volume of the sphere. This
function is discontinuous at $r=r_w$. If this is a concern, a ``mollifying
function'' \cite{murdoch2007} that smoothly takes $w(r)$ to zero at $r_w$ over some desired range can be added  (see \eref{eqn:weight_molly}).
Other possible choices include for example Gaussian
functions\cite{hardy1982}, or spline function used in meshless
methods\cite{belytchko1996} (see Ref\cite{admal2010} for details). Many physical
interpretations can be given to the weighting function. See Ref\cite{murdoch1994}
for further details. 

One possible interpretation for a positive-valued 
$w$ with compact support (as described above)
can be related to the physical nature of the experimental probe measuring
the continuum fields. In this case, the size of the compact support represents the length scale over which
continuum fields are being measured.  An alternative approach described 
by Murdoch and Bedeaux\cite{murdoch1994} is based on the
requirement that ``repeated spatial averaging should produce nothing new''.
In other words, spatially averaging a quantity that was already spatially
averaged should give the same average. This leads to a definite form for
the weighting function that also takes on negative values.

It is straightforward to see that substituting the expression given 
in \eref{eqn:w_md} for the probability density function into 
\eref{eqn:define_se_0} and performing the spatial averaging defined 
there using any weighting function discussed above,
we obtain expressions for continuum fields
that can be numerically evaluated using the data generated from an MD
simulation. For example, let us look at the mass density field given in \eref{eqn:define_density} and
repeated here with $W=W^{\rm MD}$:
\begin{align}
\rho(\bm x,t) = \sum_\alpha m_\alpha \langle W^{\rm MD} \mid \bm x_\alpha=\bm x \rangle.
\end{align}                                           
Spatially averaging this distribution with respect to the weighting function
results in 
\begin{align}
\label{eqn:density_md}
\rho_w(\bm x,t) &= \sum_\alpha m_\alpha w(\bm x_\alpha^{\rm MD}-\bm x),
\end{align}
where $\rho_w$ denotes the continuum mass density field obtained by the spatial
averaging of the pointwise field $\rho$ with respect to the weighting
function $w$. 
Similarly, all other continuum definitions for an MD simulation are obtained
from their probabilistic versions. Following is a catalog of definitions for
continuum fields that can be evaluated in any MD simulation:
\begin{widetext}
\begin{align}
\bm p_w(\bm x,t) &= \sum_{\alpha} m_\alpha \bm v_\alpha w(\bm{x}_\alpha^{\rm MD} -
\bm{x}),\\
\bm{v}_w(\bm{x},t) &= \frac{\bm{p}_w(\bm{x},t)}{\rho_w(\bm{x},t)}, \\
\stress_w(\bm x,t) &= \stress_{w,\rm k}(\bm x,t) + \stress_{w,\rm v}(\bm x,t), \\
\epsilon_w(\bm x,t) &= \epsilon_{w,\rm k}(\bm x,t) + \epsilon_{w,\rm v}(\bm
x,t), \\
\bm q_w(\bm x,t) &= \bm q_{w,\rm k}(\bm x,t) + \bm q_{w,\rm v}(\bm x,t), \\
\bm{\sigma}_{w,\rm{k}}(\bm{x},t) &= -\sum_{\alpha} m_\alpha
(\bm{v}_\alpha^{\rm{rel}} \otimes \bm{v}_\alpha ^{\rm{rel}}) w(\bm{x}_\alpha^{\rm MD} -
\bm{x}), \qquad \bm v_\alpha^{\rm rel} = \bm v_\alpha^{\rm MD}-\bm v, \\
\rho_w \epsilon_{w,\rm k}(\bm x,t) &= \frac{1}{2}\sum_\alpha m_\alpha \vnorm{\bm
v_\alpha^{\rm MD}}^2 w(\bm x_\alpha^{\rm MD} - \bm x), \\
\bm q_{w,\rm k}(\bm x,t) &= \frac{1}{2}\sum_\alpha m_\alpha \vnorm{\bm
v_\alpha^{\rm rel}}^2
\bm v_\alpha^{\rm rel} w(\bm x_\alpha^{\rm MD} - \bm x), \label{eqn:qwk_md}\\
\bm{\sigma}_{w,\rm{v}}(\bm{x},t) &= \frac{1}{2} \sum_{\substack{\alpha,\beta \\
\alpha \neq \beta}} \int_{\real{3}} w(\bm y-\bm x) 
\int_{s=0}^{1} \left \langle -\bm f_{\alpha\beta} W^{\rm MD} \mid \bm{x}_\alpha=\bm{y}+ s\bm{z},\bm{x}_\beta = \bm{y} -
(1-s)\bm{z} \right \rangle \, ds \otimes \bm z \, d\bm{z} \, d\bm y,
\label{eqn:sigmav_avg}\\
\epsilon_{w,\rm v}(\bm x, t) &= \int_0^t \frac{1}{\rho} \left ( \stress_w:\nabla_{\bm
x} \bm v_w - \bar g_{w,\rm S}(\bm x,t) \right )\, dt, 
\qquad 
\bar g_{w,\rm S}(\bm x,t) = \frac{1}{2}\sum_{\substack{\alpha,\beta \\ \alpha
\ne \beta}} \bm f_{\alpha\beta}^{\rm MD} \cdot (\bm v_\alpha^{\rm MD} - \bm
v_\beta^{\rm MD})
w(\bm x_\alpha^{\rm MD} - \bm x),
\label{eqn:ewv_md}\\
\bm q_{w,\rm v}(\bm x,t) &= \frac{1}{2} \sum_{\substack{\alpha \beta \\ \alpha \ne
\beta}} \int_{\real{3}} \!\!\! w(\bm y-\bm x)
\int_{\real{3}} \bm z \int_{s=0}^{1} \!\!
\left \langle \bm f_{\alpha\beta} \cdot \left (\frac{\bm v_\alpha \! + \!
\bm v_\beta}{2}-\bm v \right ) W^{\rm MD} \mid \bm x_\alpha = \bm y+s\bm z, \bm x_\beta
= \bm y - (1-s)\bm z \right \rangle \, ds \, d\bm z \, d\bm y. \label{eqn:qv_avg}
\end{align}
\end{widetext}
The expressions for $\stress_{w,\rm v}$ and $\bm q_{w,\rm v}$ given in
\eref{eqn:sigmav_avg}
and \eref{eqn:qv_avg} respectively,
can be further simplified by the following simple change of variables:
\begin{align}
\label{eqn:var_change_1}
\bm y + s\bm z = \bm u, \qquad \bm y-(1-s)\bm z = \bm v,
\end{align}
which implies that
\begin{align}
\label{eqn:var_change_2}
\bm z = \bm u - \bm v, \qquad \bm y = (1-s)\bm u + s\bm v.
\end{align}
The Jacobian of the transformation is
\begin{equation}
J = \det \left [ \begin{array} {cc}
\nabla_{\bm{u}} \bm{z}  & \nabla_{\bm{v}} \bm{z} \\   
\nabla_{\bm{u}} \bm{y}_{\perp}  & \nabla_{\bm{v}} \bm{y}_{\perp}
\end{array} \right ] = \det \left [ \begin{array}{cc}
  \bm{I}           & -\bm{I}       \\
(1-s) \bm{I}  & s\bm{I} 
\end{array} \right] = 1.
\label{eqn:jacobian} 
\end{equation}
Using \eref{eqn:var_change_1}, \eref{eqn:var_change_2} and
\eref{eqn:jacobian}, $\bm q_{w, \rm v}$ simplifies to
\begin{widetext}
\begin{align}
\bm q_{w,\rm v}(\bm x,t) &= \frac{1}{2} \sum_{\substack{\alpha \beta \\ \alpha \ne
\beta}} \int_{\real{3} \times \real{3}} 
\left \langle \bm f_{\alpha\beta} \cdot \left (\frac{\bm v_\alpha +
\bm v_\beta}{2}-\bm v \right ) W^{\rm MD} \mid \bm x_\alpha = \bm u , \bm x_\beta = \bm
v \right \rangle (\bm u-\bm v) b(\bm x;\bm u,\bm v) \, d\bm u \, d\bm v, \notag \\
&= \frac{1}{2} \sum_{\substack{\alpha \beta \\ \alpha \ne
\beta}} \bm f_{\alpha\beta}^{\rm MD} \cdot \left (\frac{\bm v_\alpha^{\rm MD} +
\bm v_\beta^{\rm MD}}{2}-\bm v_w \right ) (\bm x_\alpha^{\rm MD}-\bm x_\beta^{\rm
MD}) b(\bm x;\bm x_\alpha^{\rm MD},\bm x_\beta^{\rm MD}), \label{eqn:qwv_md}
\end{align}
where
\begin{align}
b(\bm x;\bm u,\bm v) = \int_{s=0}^1 w((1-s)\bm u + s\bm v - \bm x) \, ds, \notag
\end{align}
is called the \emph{bond function}. 
Similarly, $\stress_{w,{\rm v}}$ simplifies to
\begin{align}
\label{eqn:sigmawv_md}
\stress_{w,\rm{v}}(\bm{x},t) = \frac{1}{2} \sum_{\substack{\alpha,\beta \\
\alpha \neq \beta}} -\bm{f}_{\alpha\beta}^{\rm MD} \otimes (\bm{x}_\alpha^{\rm MD}-\bm{x}_\beta^{\rm MD})
b(\bm{x};\bm{x}_\alpha^{\rm MD},\bm{x}_\beta^{\rm MD}).
\end{align}
\end{widetext}
Hence, equations \eref{eqn:density_md}-\eref{eqn:qwk_md}, \eref{eqn:ewv_md}, 
\eref{eqn:qwv_md} and \eref{eqn:sigmawv_md}, are the most simplified form of continuum fields that can be evaluated in an MD simulation.
These expressions constitute a generalization to many-body
potentials of expressions originally
given by Hardy\cite{hardy1982} for the special case of pair potentials
(noting that our expressions for the internal energy density and 
heat flux vector which have modified forms). 
Other commonly-used definitions like the virial stress\cite{clausius1870}
and the Tsai traction\cite{tsai1979} can be obtained as limiting cases of the 
above relations (see Ref\cite{admal2010} for details).
We therefore see that our formulation is unified in the sense that
it shows how all of these relations are related and provides a single framework
that encompasses them all.

\section{Numerical experiments}
\label{sec:numerical}
It was mentioned at the end of \sref{sec:ener_balance} that the definitions
obtained in the new derivation are quite different and some qualitative
differences were identified. We now try to see how the definitions vary
quantitatively. Various stress tensors obtained through this
unified framework were studied in Ref\cite{admal2010} by the authors. In this
section, we describe molecular dynamics simulations that are carried out to
compare the following two quantities:
\begin{align}
\epsilon_{w,\rm v}^{\rm ik}(\bm x,t) &= \frac{1}{\rho}\sum_\alpha 
\pot_\alpha w(\bm x_\alpha^{\rm MD} - \bm x) - \epsilon_{w,\rm v}^{\rm ik}(\bm x,0), \label{eqn:ewv_ik_md_comp}\\
\epsilon_{w,\rm v}(\bm x, t) &= \int_0^t \frac{1}{\rho} \left ( \stress_w:\nabla_{\bm
x} \bm v_w - \bar g_{w,\rm S}(\bm x,t) \right )\, dt. \label{eqn:ewv_md_comp}
\end{align}
Here $\epsilon_{w,\rm v}^{\rm ik}$ is the local spatial average of the potential
part of the specific internal energy in the original Irving--Kirkwood formulation.
$\epsilon_{w,\rm v}$ is the corresponding expression in the new formulation taken
from \eref{eqn:ewv_md}.
The constant in \eref{eqn:ewv_ik_md_comp} is chosen in order to compare the
above two equations from a fixed datum. To explore the role of the two parts 
of the integrand  in
\eref{eqn:ewv_md_comp}, we define
\begin{align}
\epsilon_{w,\rm v}^1(\bm x, t) &:= \int_0^t \frac{1}{\rho} \left ( \stress_w:\nabla_{\bm
x} \bm v_w \right )\, dt,  \\
\epsilon_{w,\rm v}^2(\bm x, t) &:= -\int_0^t \frac{1}{\rho} 
\bar g_{w,\rm S}(\bm x,t) \, dt,
\end{align}
so that $\epsilon_{w,\rm v} = \epsilon_{w,\rm v}^1 + \epsilon_{w,\rm v}^2$.

\subsection*{Interatomic potential}
Since the unified framework applies to arbitrary multi-body potentials, it
would have been ideal to choose a multi-body potential for our numerical
experiments.  Unfortunately, since as mentioned in the introduction, there is no rigorous way to 
distribute the total energy among particles, the expression for $\epsilon_{w,\rm
v}^{\rm ik}$ in \eref{eqn:ewv_ik_md_comp} is ill-defined. Thus the only possibility for comparison is in
the special case of pair potential interactions in a system of identical
particles. In this case due to symmetry, it is reasonable to divide the energy
equally among the particles (see
footnote~\footnotemark[\thefnnumber]),
thus arriving at a plausible definition for $\epsilon_{w,\rm v}^{\rm ik}$. 
We will see that even in this case the expressions in \eref{eqn:ewv_ik_md_comp} 
and \eref{eqn:ewv_md_comp} are not equivalent. In more general cases, we argue
that only the new expression in \eref{eqn:ewv_md_comp} is well-defined.

Our system consists of particles arranged in a face-centered cubic crystal, stacked together in
the form of $15\times15\times15$ unit cells, interacting through
a modified Lennard-Jones type potential.  The Lennard-Jones parameter,
$\epsilon$ and $\sigma$ are arbitrarily set to 1.  The potential has the following form:
\begin{align}
\label{eqn:lennard}
\phi(r) = 4 \left [ \frac{1}{r^{12}} - \frac{1}{r^6} \right ] - 0.0078 r^2 + 0.0651.
\end{align}
The above equation has been rendered dimensionless by expressing lengths in
units of $\sigma$ and energy in units of $\epsilon$. As seen in the above
equation, the standard Lennard-Jones potential is modified by the addition of a
quadratic term. This is done to ensure that 
$\phi(r_{\rm cut}) = \phi'(r_{\rm cut}) = 0$, where
$r_{\rm cut} = 2.5$, denotes the cutoff radius for the potential. We refer to
this as the ``Modified Lennard-Jones potential''. The ground state of this
potential is an fcc crystal with a lattice constant of $a=1.556$. Thus, the
dimension of the sample at ground state is given by its length
$l=15\times1.556=23.34$. We use the
\emph{velocity Verlet} time integration algorithm to evolve the system.

The weighting function, $w(r)$,
is chosen to be a constant with a suitable mollifying function,\cite{murdoch2007}
\begin{equation}
\label{eqn:weight_molly}
w(r) = \left \{ \begin{array}{ll}
c & \mbox{if $r<R-\delta$}\\
\frac{1}{2}c\left[ 1- \cos \left( \frac{R-r}{\delta}\pi \right ) \right ] &
\mbox{if $R-\delta<r<R$} \\
0 & \mbox{otherwise} \end{array} \right.,
\end{equation}
where $\delta = 0.12$, $R$ is the radius of the sphere which forms the compact
support and $c = 1 - (\delta / R)^3 + 3(\delta / R)^2 - 1.5
(\delta/ R)$. The constant $c$ is chosen to normalize the weighting function. 

\subsection*{Experiment 1}
\begin{figure*}
\centering
\includegraphics[scale=0.7]{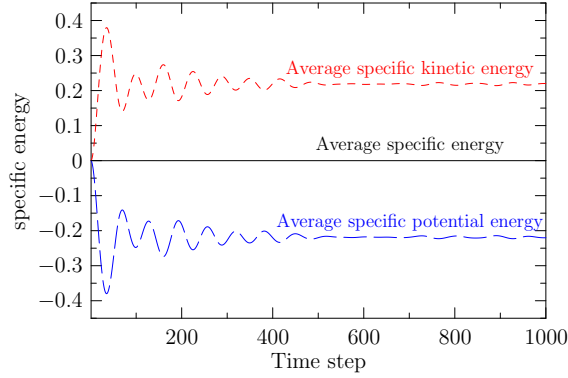}
\caption{Plot showing the average specific potential energy and the average
specific kinetic energy. Since this is a constant energy simulation, their sum
(\emph{black solid line}) is always constant.}
\label{fig:pbc_e_total}
\end{figure*}
\begin{figure*}
\centering
\subfigure[]{\includegraphics[scale=0.7]{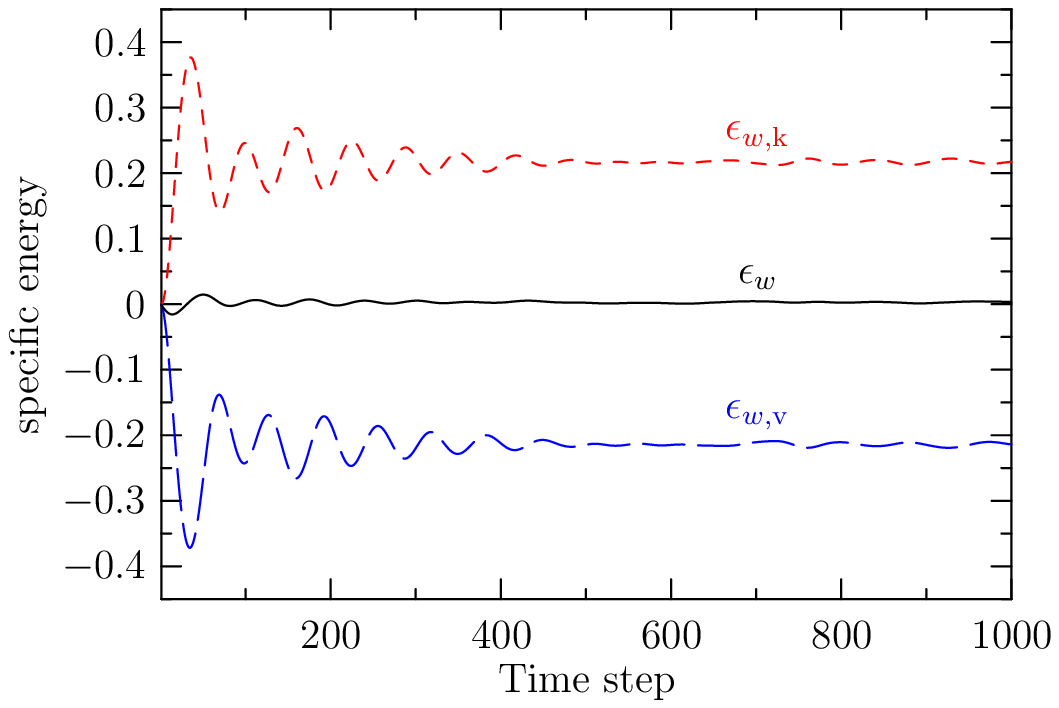} \label{fig:pbc_e}}
\subfigure[]{\includegraphics[scale=0.7]{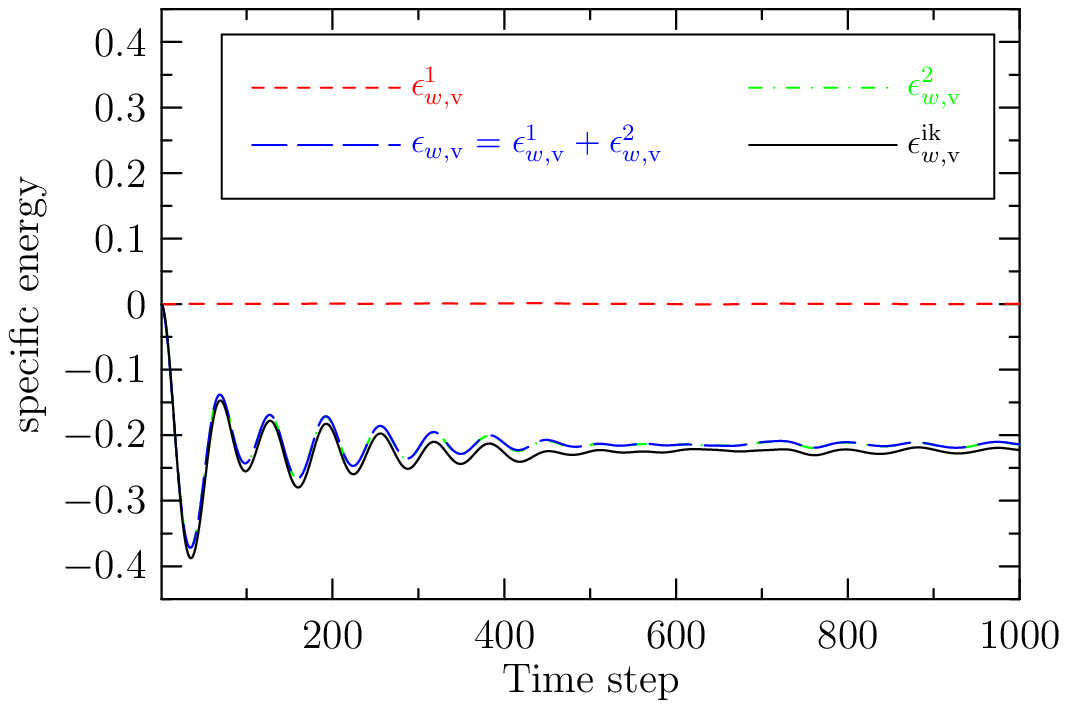}  \label{fig:pbc_ev_compare}}
\caption{Evolution of specific internal energy for a constant energy simulation with
periodic boundary conditions. 
(a) Plot showing the evolution of the potential part, $\epsilon_{w,\rm
v}$, and the kinetic part, $\epsilon_{w,\rm k}$, of the specific internal energy.  The
total specific internal energy (shown in black solid line) is not strictly constant. 
(b) Plot comparing the evolution of the potential part of
the specific internal energy with its analogue, $\epsilon_{w,\rm v}^{\rm ik}$, in the
original Irving--Kirkwood procedure.}
\label{fig:pbc_constant_e}
\end{figure*}
We begin with a \emph{constant energy} molecular dynamics simulation with \emph{periodic boundary
conditions}.  The atoms in the system are randomly perturbed to induce a temperature of $T=0.145$
in Lennard-Jones units after equilibration. \fref{fig:pbc_e_total} shows the evolution of
the average specific potential energy, i.e., the total potential energy divided
by the mass ($=N$ in our case), and average specific kinetic energy,
which add up to a constant specific internal energy. (The average specific
internal energy is constant since this is a constant energy simulation.)

Now, suppose we are interested in evaluating the continuum quantities given in
\eref{eqn:ewv_ik_md_comp} and \eref{eqn:ewv_md_comp} at the
center of our system. To do this, we choose $R=0.4l$, which corresponds to a
length of $6$ unit cells, for the radius of the
compact support of the weighting function. 

\fref{fig:pbc_e} shows a plot of the potential and kinetic part of the specific
energy at the center of the sample calculated using the weighting function given
in \eref{eqn:weight_molly}. It is clear from the plot that the total specific
energy at the center of the sample has some oscillations up to about $200$ time
steps before these oscillations become negligible. This reminds us that 
although we are performing a constant energy simulation, the \emph{specific} internal energy at a point need not be constant. 
\fref{fig:pbc_ev_compare} compares the different expressions
for the potential part of the specific internal energy given in \eref{eqn:ewv_ik_md_comp}
and \eref{eqn:ewv_md_comp}. The plots for $\epsilon_{w,\rm v}$ and
$\epsilon_{w,\rm v}^{\rm ik}$ are in good agreement. It is also clear from the plot that the
contribution of $\epsilon_{w,\rm v}^1$ to $\epsilon_{w,\rm v}$ is
negligible.

\subsection*{Experiment 2}
\begin{figure*}
\centering
\subfigure[]{\includegraphics[scale=0.7]{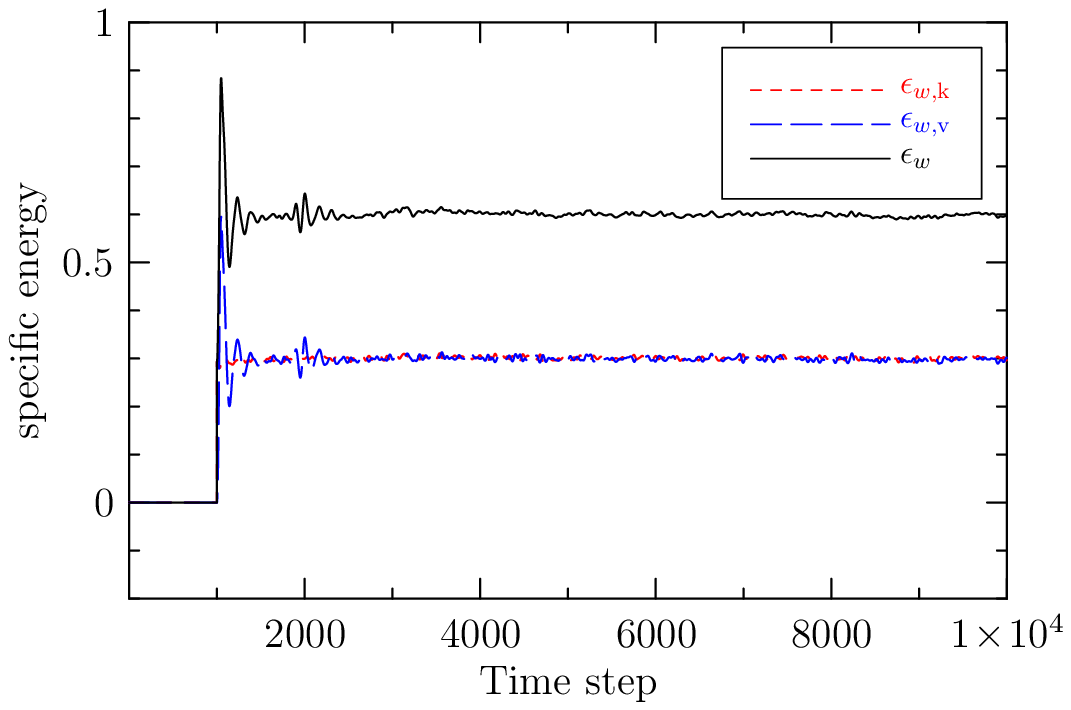} \label{fig:pbc_rsp0.8_e}}
\subfigure[]{\includegraphics[scale=0.7]{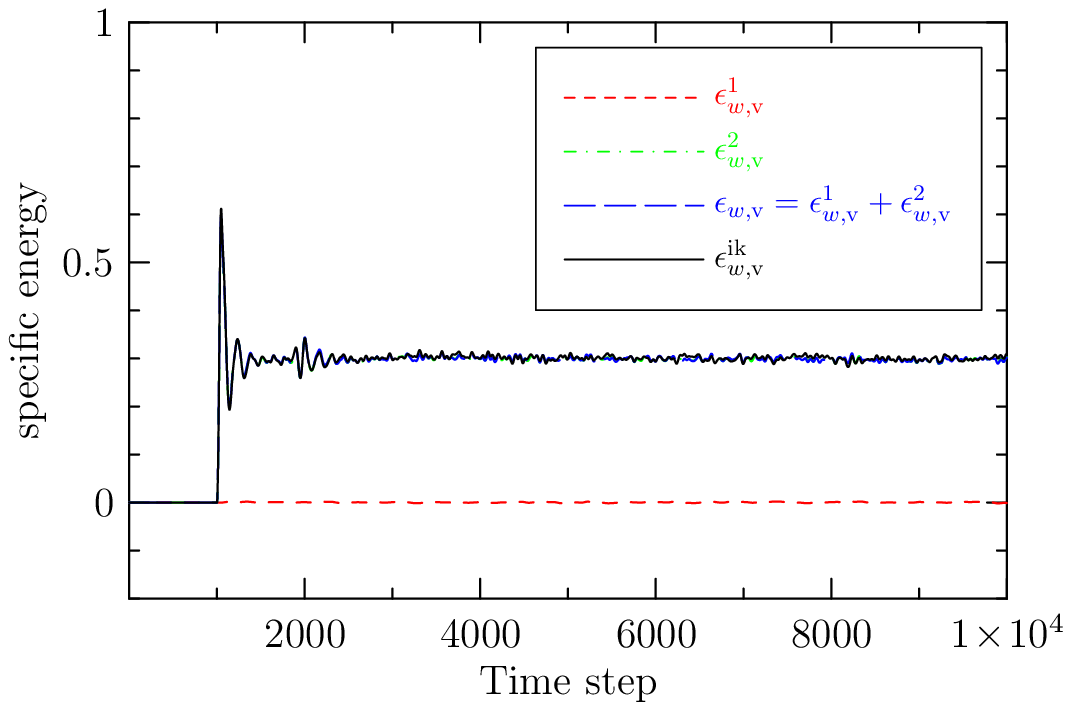} \label{fig:pbc_rsp0.8_ev_compare}}
\caption{Evolution of specific internal energy for a constant temperature (applied after
first $1000$ time steps) simulation with periodic boundary conditions. 
(a) Plot showing the evolution of the potential part, $\epsilon_{w,\rm
v}$, and the kinetic part, $\epsilon_{w,\rm k}$, of the specific internal energy 
(b) Plot comparing the evolution of the potential part of
the specific internal energy with its analogue, $\epsilon_{w,\rm v}^{\rm ik}$, in the
original Irving--Kirkwood procedure.}
\label{fig:pbc_rsp0.8}
\end{figure*}
In this experiment, we continue with the same system with periodic boundary
conditions. We begin with the particles at their equilibrium
positions, with temperature $T=0$. We allow the system to evolve for the first $1000$
time steps during which we observe small fluctuations due to numerical noise.
We then instantaneously increase the temperature to $T=0.2$ using a simple
velocity rescaling thermostat and maintain this temperature for the rest of the
simulation. Again, we are interested in studying the continuum fields at the
center of the sample.  We choose the radius of the compact support $R=0.4l$ for
the weighting function, as in the previous experiment.
\fref{fig:pbc_rsp0.8_e} shows the plot of $\epsilon_{w,\rm
v}$, $\epsilon_{w,\rm k}$, and the total specific internal energy at the center of the sample for this case. \fref{fig:pbc_rsp0.8_ev_compare} compares
$\epsilon_{w,\rm v}$ with $\epsilon_{w,\rm v}^{\rm ik}$. It is clear from this
plot that both $\epsilon_{w,\rm v}$ and $\epsilon_{w,\rm v}^{\rm ik}$ are in
good agreement with each other. It is interesting to see that the contribution
to $\epsilon_{w,\rm v}$ due to $\epsilon_{w,\rm v}^1$ remains negligible even after
increasing the temperature of the system.

\subsection*{Experiment 3}
\begin{figure*}
\centering
\subfigure[]{\includegraphics[scale=0.7]{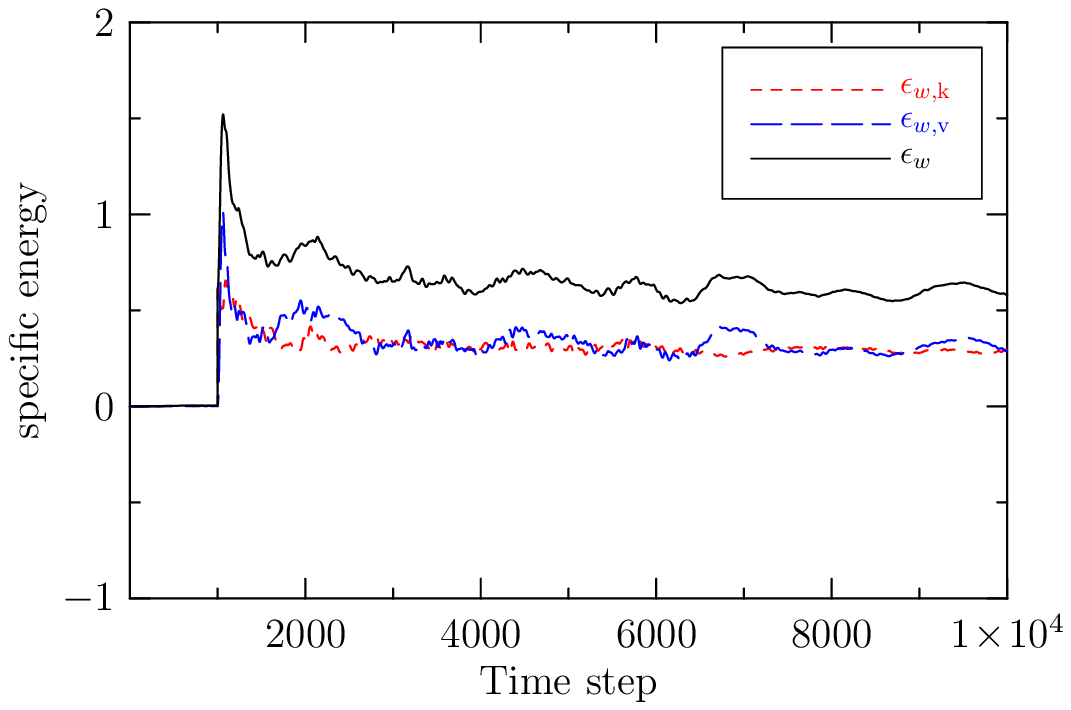} \label{fig:nopbc_rsp0.8_e}}
\subfigure[]{\includegraphics[scale=0.7]{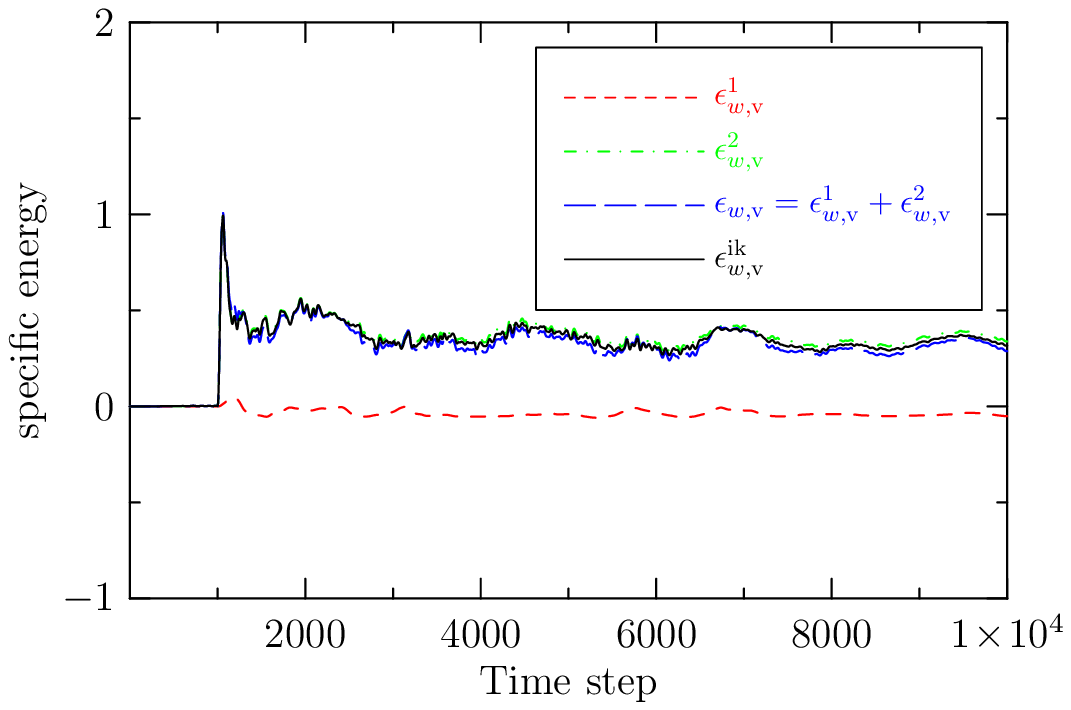} \label{fig:nopbc_rsp0.8_ev_compare}}
\caption{Evolution of specific internal energy for a constant temperature (applied after
first $1000$ time steps) simulation \emph{without} periodic boundary conditions,
using an averaging domain of radius $R=0.4l$.
(a) Plot showing the evolution of the potential part, $\epsilon_{w,\rm
v}$, and the kinetic part, $\epsilon_{w,\rm k}$, of the specific internal energy. 
(b) Plot comparing the evolution of the potential part of the specific internal energy
with its analogue, $\epsilon_{w,\rm v}^{\rm ik}$, in the original Irving--Kirkwood
procedure.}
\label{fig:nopbc_rsp0.8}
\end{figure*}
\begin{figure*}
\centering
\subfigure[]{\includegraphics[scale=0.7]{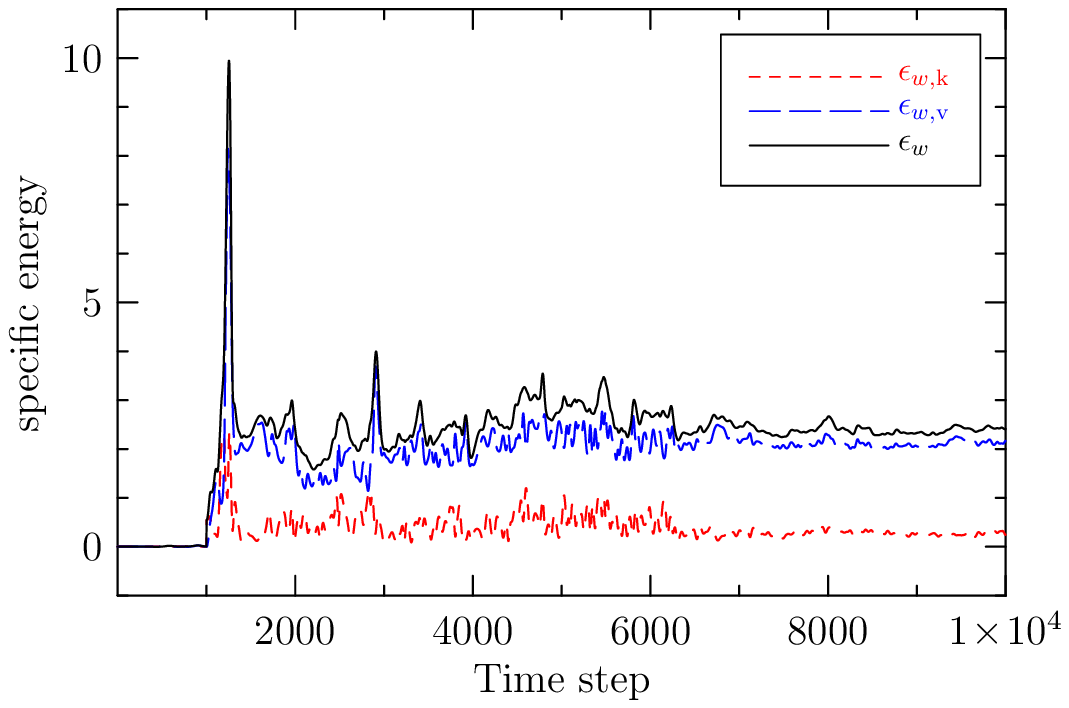} \label{fig:nopbc_rsp0.2_e}}
\subfigure[]{\includegraphics[scale=0.7]{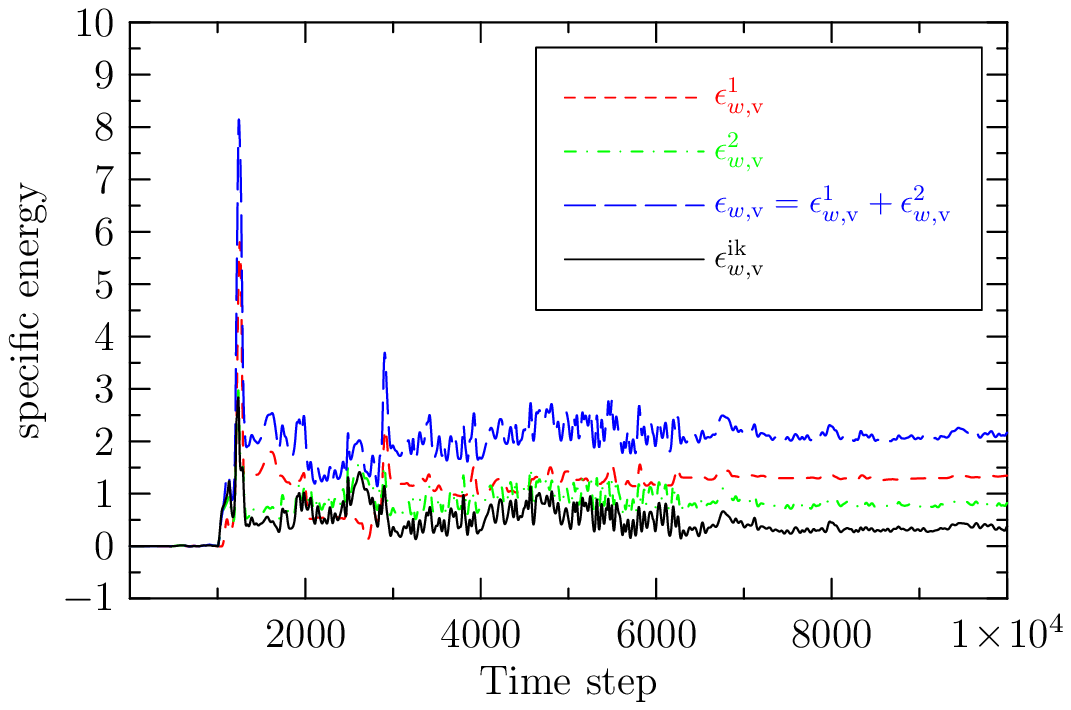} \label{fig:nopbc_rsp0.2_ev_compare}}
\caption{Evolution of specific internal energy for a constant temperature (applied after
first $1000$ time steps) simulation \emph{without} periodic boundary conditions,
using an averaging domain of radius $R=0.1l$.
(a) Plot showing the evolution of the potential part, $\epsilon_{w,\rm
v}$, and the kinetic part, $\epsilon_{w,\rm k}$, of the specific internal energy. 
(b) Plot comparing the evolution of the potential part of the specific internal energy
with its analogue, $\epsilon_{w,\rm v}^{\rm ik}$, in the original Irving--Kirkwood
procedure.}
\label{fig:nopbc_rsp0.2}
\end{figure*}
The setup is similar to the previous experiment except that now we do not apply
periodic boundary conditions. This means that the sample is free to expand once
the temperature of the sample is increased. The plots shown in
\fref{fig:nopbc_rsp0.8} correspond to $R=0.4l$. \fref{fig:nopbc_rsp0.8_e} shows the
plot of $\epsilon_{w,\rm v}$, $\epsilon_{w,\rm k}$ and the total specific
energy at the center of the sample, and \fref{fig:nopbc_rsp0.8_ev_compare} compares
$\epsilon_{w,\rm v}$ with $\epsilon_{w,\rm v}^{\rm ik}$. In this case we see
from \fref{fig:nopbc_rsp0.8_ev_compare} that unlike the previous two experiments
there is a small negative contribution to $\epsilon_{w,\rm v}$ due to $\epsilon_{w,\rm v}^1$ --- although the magnitude is still very small.
Note that the longer wavelength oscillations in energy in these plots 
correspond to the pulsing of the sample as it expands and contracts about 
its mean thermally-expanded size.

It is also interesting to see how these continuum fields change as the averaging
domain size is decreased. As mentioned previously, as the weighting function tends
to the Dirac delta distribution, we expect the continuum fields to also become localized
with increasing magnitude.
To see if this is indeed the case, we decrease
the size of the averaging domain by choosing $R=0.1l$, which corresponds to a
length of $1.5$ unit cells. \fref{fig:nopbc_rsp0.2}
shows the plots for this case. Comparing \fref{fig:nopbc_rsp0.2_e} with 
\fref{fig:nopbc_rsp0.8_e}, we see that $\epsilon_{w,\rm k}$ remains the same, while
$\epsilon_{w,\rm v}$ for the smaller averaging domain is about four times larger than 
its value for the larger domain. Similarly comparing \fref{fig:nopbc_rsp0.2_ev_compare}
with \fref{fig:nopbc_rsp0.8_ev_compare}, we see that for the smaller domain,
$\epsilon_{w,\rm v}$ is greater than four times
$\epsilon_{w,\rm v}^{\rm ik}$, whereas they were equal for the larger domain. 
Moreover, the contribution due to
$\epsilon_{w,\rm v}^1$ to the specific internal energy is larger than that due to
$\epsilon_{w,\rm v}^2$. This clearly shows that for small averaging domains, the
two definitions given by \eref{eqn:ewv_ik_md_comp} and \eref{eqn:ewv_md_comp}
are quite different in nature. Based on our observations, we can conclude that 
$\epsilon_{w,\rm v}$ tends to localize more strongly with the averaging domain size than does $\epsilon_{w,\rm v}^{\rm ik}$. 

To rationalize these results, let us consider what happens to these definitions
as the weighting function tends to a delta distribution.
In this case, $\epsilon_{w,\rm v}^{\rm ik}$ (see \eref{eqn:ewv_ik_md_comp})
localizes to the particle positions. The same also happens to all terms in 
the definition of $\epsilon_{w,\rm v}$ (see \eref{eqn:ewv_md_comp}) 
\emph{except} for $\sigma_w$. This term localizes to the lines joining the
particles rather than onto the particles themselves.
This observation provides a qualitative explanation, although not complete, for the different
behavior of the two definitions shown above.

\section{Summary}
In this paper, we present a two-step \emph{unified framework} for the evaluation of 
continuum field expressions from molecular simulations: (1) pointwise continuum
fields are obtained using a generalization of the Irving--Kirkwood procedure
to arbitrary multi-body potentials, and (2) spatial averaging is applied to obtain
the corresponding macroscopic quantities. The latter lead to expressions suitable
for computation in molecular dynamics simulations. It is shown that the important
commonly-used microscopic definitions for continuum fields are recovered in this
process as special cases.

In generalizing the Irving--Kirkwood to arbitrary many-body potentials we
have had to address two ambiguities inherent in the original procedure which
lead to non-uniqueness in the resulting expressions.
The first ambiguity arises due to the non-uniqueness of the
partitioning of the force on an atom as a sum of central forces,
which is directly related to the non-uniqueness of the
potential energy representation in terms of distances between particles.
This is in turn related to the shape space of the system. The conclusion
is that the pointwise stress tensor is \emph{not} unique, however
we show in Ref\cite{admal2010}
that the macroscopic stress obtained via spatial averaging becomes
unique as the spatial averaging domain is taken to infinity.

The second ambiguity in the original Irving--Kirkwood procedure
arises due to the arbitrary decomposition of energy between
particles. We show that this decomposition can be completely avoided through
an alternative derivation for the energy balance equation. This leads to
new definitions for the specific internal energy and the heat flux
vector. In particular, the resulting potential part of the specific internal energy
does not depend on the arbitrary partitioning of the potential energy to
individual particles, and the resulting heat flux vector does not contain the ``transport part''
which is not invariant with respect to the addition of a constant to the 
potential energy function.

The new definition for the specific internal energy is compared with
the original Irving--Kirkwood definition through a series of numerical experiments.
Although our expression applies to arbitrary many-body potentials, we have chosen 
to perform the comparisons for the special case of a system of identical particles
interacting through a pair potential since this is the only case where the original 
Irving--Kirkwood internal energy density is well-defined.
This is due to the ambiguity in the decomposition of energy between the particles
in the original Irving--Kirkwood derivation (and existing extensions to the approach).  
In our numerical experiments, we observe that both definitions
agree for weighting functions with support
given by a ball of radius 0.6 unit cells and larger. However, as the
weighting function tends to a delta distribution, the two definitions scale
differently. A qualitative theoretical explanation for this difference is given based 
on the limiting behavior of the two definitions as the averaging domain
tends to a point.

It would also be of interest to compare the new expression for the heat 
flux vector to the original Irving--Kirkwood expression.
In order to do this, one has to study the transport part of the
heat flux. To our knowledge,
this has not been done in the past because the transport part is normally
lumped into either the kinetic or potential parts of the heat flux and not 
observed separately.
Our new derivation shows that its
existence is directly related to the arbitrariness in the energy decomposition.
Preliminary numerical studies of some systems (not reported here) 
suggest that the transport part of the heat flux vector, which is absent in 
the new derivation, tends to be negligible. Further work is
necessary to determine whether this is a general result.

\begin{acknowledgments}
This research was partly supported through the National Science
Foundation (NSF) under Grant No. DMS-0757355 and the Air Force Office of
Scientific Research (AFOSR) under Grant No. FA9550-09-1-0157. The views
and conclusions contained herein are those of the authors and should not
be interpreted as necessarily representing the official policies or
endorsements, either expressed or implied, of the AFOSR or the U.S.
Government.
\end{acknowledgments}
 

%

\end{document}